\documentclass[preprint,5p,times]{elsarticle} 

\usepackage{amsmath,amssymb,hyperref}
\usepackage{xcolor}
\usepackage{bm}
\usepackage{wrapfig}
\usepackage{changepage}
\usepackage{ulem}
\usepackage{float}
\usepackage{algorithm}
\usepackage{algpseudocode}

\newtheorem{rmk}{Remark}
\usepackage{tikz}
\usetikzlibrary{arrows.meta, 
                chains,
                ext.paths.ortho,    
                positioning,
                shapes.geometric}

\tikzset{FlowChart/.style={
     base/.style = {rectangle, draw,
                    minimum width=22mm, minimum height=1cm, align=flush center,
                    },
startstop/.style = {base},
  process/.style = {base, 
                    text width=5cm},
 decision/.style = {diamond, draw, sharp corners, aspect=1.3, 
                    text width=3cm, inner sep=0mm, align=flush center},
    arrow/.style = {thick,-Triangle},
        }   }
\makeatletter
\tikzset{suspend join/.code={\def\tikz@after@path{}}}
\makeatother
\usetikzlibrary{arrows,shapes,calc}
\usepackage{pgfplots}
\usepackage{enumitem}  
\usetikzlibrary{positioning}
\tikzstyle{solid}=                   [dash pattern=]
\tikzstyle{dotted}=                  [dash pattern=on \pgflinewidth off 2pt]
\tikzstyle{densely dotted}=          [dash pattern=on \pgflinewidth off 1pt]
\tikzstyle{loosely dotted}=          [dash pattern=on \pgflinewidth off 4pt]
\tikzstyle{dashed}=                  [dash pattern=on 3pt off 3pt]
\tikzstyle{densely dashed}=          [dash pattern=on 3pt off 2pt]
\tikzstyle{loosely dashed}=          [dash pattern=on 3pt off 6pt]
\tikzstyle{dashdotted}=              [dash pattern=on 3pt off 2pt on \the\pgflinewidth off 2pt]
\tikzstyle{densely dashdotted}=      [dash pattern=on 3pt off 1pt on \the\pgflinewidth off 1pt]
\tikzstyle{loosely dashdotted}=      [dash pattern=on 3pt off 4pt on \the\pgflinewidth off 4pt]
\usepackage[skins,breakable,many]{tcolorbox}

\tcbsetforeverylayer{
  fonttitle=\bfseries,
  left=.1\textwidth,
  right=.1\textwidth,
  breakable,
  enhanced,
  interior hidden,
  frame hidden,
  coltitle=black,
}

\newcommand{\e}{\vskip 2mm\noindent}

\usepackage{graphicx}      
\newtcolorbox{rightbrace}{%
    enhanced jigsaw, 
    breakable, 
    frame hidden, 
    overlay={%
        \draw [
            fill=none, 
            decoration={brace,amplitude=0.5em},
            decorate,
            ultra thick,
            gray,
        ]
        (frame.north east)--(frame.south east);
    },
    parbox=false,
}
\begin{document}
\begin{frontmatter}

\title{Model-Free Unsupervised Anomaly Detection Framework \\ in  Multivariate Time-Series of Industrial Dynamical Systems}
\journal{International Journal of Control}
\author[gipsa]{Mazen Alamir \corref{cor1}}\ead{mazen.alamir@grenoble-inp.fr}
\author[gipsa]{Rapha\"el Dion\corref{cor2}}\ead{raphael.dion@grenoble-inp.fr}
\address[gipsa]{Univ. Grenoble Alpes, CNRS, Grenoble INP, GIPSA-lab, 38000 Grenoble, France.}
\cortext[cor2]{Corresponding author}


\begin{abstract}                          
In this paper, a new model-free anomaly detection framework is proposed for time-series induced by industrial dynamical systems. The framework lies in the category of conventional approaches which enable appealing features such as a learning with reduced amount of training data, a high potential for explainability as well as a compatibility with incremental learning mechanism to incorporate operator feedback after an alarm is raised and analyzed. Although these are crucial features towards acceptance of data-driven solutions by industry, they are rarely considered in the comparisons that generally almost  exclusively focus on performance metrics. Moreover, the features engineering step involved in the proposed framework is inspired by the time-series being implicitly governed by physical laws as it is generally the case in industrial time-series. Two examples are given to assess the efficiency of the proposed approach.
\end{abstract}
\end{frontmatter}

\section{Introduction}\label{sec_intro}
Model-free anomaly detection in industrial time-series is a major challenge for obvious safety and economic reasons \cite{Raheem2022}. This challenge is even harder to tackle in the unsupervised (or one-class \cite{Moya1993}) setting, namely, when the model used in the decision process is to be built on healthy time-series and in the absence of any mathematical knowledge-based model \cite{Pota2023}. \\ \\
It is a fact that this setting is the only realistic one in many industrial contexts where no reliable mathematical models are available and given the scarcity of \textit{labelled} anomalous instances that can hardly cover all possible deviations from normality \cite{Wu2020}. The absence of a mathematical model governing the industrial time-series excludes the design of dynamic observers. This is the reason why observer-based anomaly detection methodologies are not analyzed in the present contribution. \\ \ \\
To be more specific, the paradigm of model-free unsupervised anomaly detection in time-series can be stated as follows: 
\begin{center}
\begin{tikzpicture}
\node[rounded corners, fill=black!10, inner sep=4mm](P){
\begin{minipage}{0.45\textwidth}Use sensors recordings containing healthy set of time-series to design an anomaly detector that takes as argument a new set of time-series (of given length) and produce a diagnosis regarding whether this set should be considered as normal or anomalous (novel).
\end{minipage} 
};
\end{tikzpicture}
\end{center} 
The contribution of this work is multiple, namely: \\ \ \\
1) A novel model-free algorithm with explainable results for fault detection in industrial contexts is proposed, with a physical laws-inspired framework.\\ \ \\
2) An Incremental Learning\cite{Souiden2022} extension of the algorithm is introduced in order to accommodate for operators' feedback.\\ \ \\
3) Two new synthetic datasets for fault detection are presented, simulating realistic dynamic systems to serve as benchmark.\\ \ \\
4) A comparison is proposed with commonly used fault detection algorithms together with an analysis regarding the influence of the solution's parameters.
\\ \ \\
The remainder of this paper is organized as follows: Section \ref{sec_background} presents an overview of the various approaches for Anomaly Detection based on which Section \ref{sec_pos_algo} positions the proposed algorithm. The latter is presented in Section \ref{sec_algo}. Section \ref{sec_examples} introduces two new synthetic datasets used in the comparison of Section \ref{sec_bench}. Finally, Section \ref{sec_conclusion} concludes the paper and proposes some tracks for future investigation. 
\section{Background}\label{sec_background}
An \textbf{Anomaly Detection} algorithm is used to discriminate between normal behavior conforming data and outliers (or novelties). It can be applied on various data shapes, such as spatial points, images or time-series. Algorithm can be derived from different types of training data ranging from fully labeled ones (\textit{supervised}) to unlabeled ones (\textit{unsupervised}) but also half-labeled sets (\textit{semi-supervised}, \textit{one-class}). The latter is the one considered in this paper.  \\ \\
This topic has attracted a huge amount of work in the last recent years fired the need for many survey papers providing a comprehensive categorization (see \cite{Blazquez2021, Chandola2009, Shaukat2021, Takeishi2014} and the references therein). This is briefly sketched in the next section.
\subsection{State-of-the-art}\label{subsec_sota}
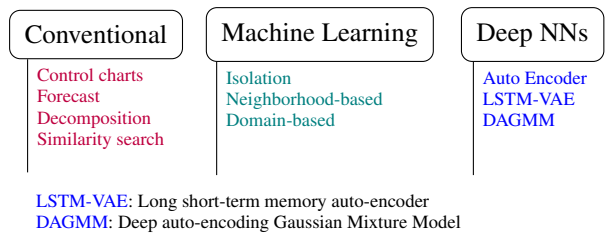
\begin{figure}[h]
\begin{tikzpicture}
\node[draw, rounded corners, inner sep=2mm] at(-3,-1)(C){Conventional};
\node[draw, rounded corners, inner sep=2mm] at($(C.east)+(1.8,0)$)(ML){Machine Learning};
\node[draw, rounded corners, inner sep=2mm,] at($(ML.east)+(1.3,0)$)(DNN){Deep NNs};
\draw (C.200) coordinate (R1) -- +(0,-1.5);
\node[anchor=north west] at(R1){\footnotesize \color{purple}
\begin{minipage}{0.33\textwidth}
Control-charts \\
Forecast \\
Decomposition\\
Similarity search
\end{minipage} 
};
\draw (ML.195) coordinate (R2) -- +(0,-1.5);
\node[anchor=north west] at(R2){\footnotesize \color{teal}
\begin{minipage}{0.33\textwidth}
Isolation \\
Neighborhood-based \\
Domain-based
\end{minipage} 
};
\draw (DNN.205) coordinate (R3) -- +(0,-1.5);
\node[anchor=north west] at(R3){\footnotesize \color{blue}
\begin{minipage}{0.33\textwidth}
Auto Encoder \\
LSTM-VAE \\
DAGMM
\end{minipage} 
};
\node[anchor=north west, below] at(-1,-3){
\scriptsize
\begin{minipage}{0.4\textwidth}
{\color{blue}LSTM-VAE}: Long Short-Term Memory Variational Auto-Encoder\\
{\color{blue}DAGMM}: Deep Auto-encoding Gaussian Mixture Model	
\end{minipage} 
};
\end{tikzpicture}
\caption{Classification of anomaly detection methods in time-series as suggested by \cite{Audibert2022}.}\label{fig_algo_categories}	
\end{figure}

\noindent According to the excellent survey \cite{Audibert2022}, the existing frameworks can be divided into three categories: \e
1) \textbf{Conventional} methods, \\ 
2) \textbf{Machine Learning}-based methods and \\ 
3) \textbf{Deep Neural Networks}-based methods. \\ \ \\ 
Figure \ref{fig_algo_categories}) shows the sub-categories mentioned by \cite{Audibert2022} within each of the above frameworks.
In the following section, these categories are briefly described in order to support the discussion of Section \ref{subsec_discussions}. Notice that describing a design approach for anomaly detection amounts at explaining \textbf{how it computes \textit{residuals} from \textit{raw sensors data}}. The residual is then compared to an elaborated threshold in order to transform the real value into a binary decision. Note also that the way to appropriately determining the threshold once the residual is generated is a crucial step on its own, which is common to all the approaches and which is not discussed in the present paper although further comments on this issue are provided in Section \ref{sec_bench} and in the conclusion.
\subsubsection{Conventional methods}\label{subsec_conventional}
\noindent These methods are denoted as \textit{conventional} since they do not involve any Machine Learning step, and they can be split into four categories. \\ \\
In the so-called \textbf{control-chart}-like approaches, auxiliary variables are generated, using the learning data, by integrating \cite{Woodall1985} or filtering \cite{Lowry1992} the excursions (away from the mean values) of some multivariate statistics. Then by characterizing bounds on the resulting excursions, the \textit{normal} range of variations is determined. The residual is then defined to be the distance to the so computed \textit{normality region}. \\ \\
In the \textbf{forecasting}-based methods, the residual is taken to be the prediction error based on a (generally linear) auto-regressive model that takes the previous values of the sensors as inputs and attempts to predict the next values \cite{Loredo2002}. Notice that as explained later (Section \ref{subsec_dnn}), the family of auto-encoder-based methods also uses the prediction errors while using DNNs as the underlying prediction model. \\ \\
The normality in the \textbf{decomposition} methods is defined by the persistent relevance of a reduced dimensional representation of the time-series using different dimensionality reduction paradigms such as \textit{Principal Component Analysis} \cite{Takeishi2014}, \textit{Singular Spectrum Analysis} \cite{Zhigljavsky2010} capturing the main dynamic-related correlations or \textit{Independent Component Analysis} \cite{Reza2015} that are intended to separate independent processes governing the dynamics. In all these cases, the residual is defined by the reconstruction errors of the signal when projected on the lower dimensional subspace. \\ \\
Finally, in the \textbf{similarity} search approaches, the sub-sequences of a time-series are described by matrices of indicators (patterns) and the residuals are defined to be the minimum distances to the elements of a set of matrices representing the normality in the training data \cite{Ren2017,ChinChia2016}. \\ 
\subsubsection{Machine Learning-based methods}\label{subsec_machine_learning}
\noindent These methods share the fact that at some stage, a Machine Learning tool is used to characterize a sample (that can be a window of raw time-series) as normal or as an outlier. As suggested by \cite{Domingues2018}, this class of methods can be split into three categories. \\ \\
In the \textbf{isolation}-based methods, decision trees are used \cite{Liu2008} in order to attempt to isolate the sample from the rest of the data. The lower the number of \textit{cutoff} operations that are needed before isolation is obtained, the higher is the probability that the sample is anomalous. \\ \\
In the \textbf{neighborhood}-based methods, an anomalous measurements is characterized by the low density of its neighbors determined via clustering methods such as \textit{K-Nearest neighbors} \cite{Breunig2000}, or density-based spatial clustering of applications with noise (\textit{DBSCAN}) \cite{Ester1996}. \\ \\
Finally, in the \textbf{domain}-based approaches, the normality domain (hyper-spheres/cubes to cite but two examples) is computed using \textit{One-Class Support Vector Machine} (OC-SVM) and the distance to the resulting domain is considered to be the residual that fires anomalous diagnosis flag when it goes beyond the computed threshold \cite{Scholkopf2001}. \\ \\
Notice that these approaches do not exploit the specificity of time-series as they consider a times-series over a window of length $n$ as a point in $\mathbb{R}^n$. Here, non time-series-specific methods are somehow \textit{recycled} to handle time-series as a particular example. Nevertheless, there exist a few Machine Learning-based algorithms that are time-aware, such as \textit{Hidden Markow Models} (HMM) or time-neighbors clustering using \textit{Dynamic Time Warping} (DTW) \cite{Tormene2009}. As the computing time of these approaches explodes with a large number of sensors or observations, this work won't elaborate further on them. \\
\subsubsection{Deep Neural Networks-based methods}\label{subsec_dnn}
\noindent These methods can be grouped into a set of increasingly sophisticated versions of the basic \textbf{Auto-Encoder} (AE) \cite{Rumelhart1985} where a specific architecture of DNN is used involving an intermediate low density layer performing a kind of dimensionality reduction task. The ability to reconstruct, with a good precision, the raw time-series by the remaining part of the DNN out of the resulting \textit{relatively smaller number of features} is a sign of normality\footnote{This recalls the principle of the decomposition methods mentioned in Section \ref{subsec_conventional}.}. The residual is therefore defined as the reconstruction error. \\ \\ The \textbf{UnSupervised Anomaly Detector} (USAD) is a more sophisticated version of the AE in which two AEs are used to play adversarial training game (one of the AEs draw artificial samples to fool the other one) \cite{Audibert2020}. This enables to enrich the training data while using the same amount of real measurements.  \\ \\
The normality can also be associated to the persistent relevance of some time dependencies characterizing the normal time-series. This is exploited in the so-called \textbf{Long Short-Term Memory Variational Auto-Encoder} (LSTM-VAE) \cite{Park2018}. \\ \\
A recent version of the Auto-Encoder principle is proposed \cite{Zong2018} where the low dimensional compressed intermediate layer is fed into a \textbf{Gaussian Mixture Model} (GMM) to build the reconstruction model. The parameters of both the AE and the GMM are optimized in order to reduce the reconstruction error which is used as residual. 
\subsection{Discussions}\label{subsec_discussions}
\subsubsection{On the No Free Lunch theorem} \label{subsec_nfl_theorem} 
\noindent In their excellent contribution \cite{Audibert2022}, the authors showed that when comparing up to sixteen anomaly detectors lying in the categories surveyed in Section \ref{subsec_sota} over five known and widely used benchmark problems, a new validation of the \textbf{No Free Lunch} theorem \cite{Stavros2019} clearly appears, namely:
\begin{center}
\begin{tikzpicture}
\node[rounded corners, fill=black!10, inner sep=4mm](P){
\begin{minipage}{0.45\textwidth}No family of method \textbf{systematically} outperforms the others \cite{Audibert2022}.
\end{minipage} 
};
\end{tikzpicture}
\end{center}
It also clearly appears that while the DNNs-based approaches might slightly better capture some dependencies that are difficult to grasp by the other tested methods, not surprisingly, the performance of theses approaches drop quickly when removing a part of the training data. In particular, as far as the involved benchmark problems are concerned, it appears that:
\begin{center}
\begin{tikzpicture}
\node[rounded corners, fill=black!10, inner sep=4mm](P){
\begin{minipage}{0.45\textwidth}Conventional methods globally obtain better results when the dataset size is less than 50\% of the original one \cite{Audibert2022}.
\end{minipage} 
};
\end{tikzpicture}
\end{center}
One by-product of the comparison done by \cite{Audibert2022} is that ML-based solutions seem to be less appropriate to handle time-series specificities. This is not surprising given that these algorithms consider, most of the time, time-series instances as simple points in a high dimensional space. \\ \\
On the other hand, DNN-based approaches suffer from the \textbf{explainability} issue, which is of a tremendous importance in the industrial context we are focusing on in the present paper. Moreover, it happens that the lack of data is even more critical in this context because of the variety of operational conditions, which renders the task of separating anomalies from unseen normal contexts of use even harder without a huge amount of data, should over-parameterized underlying models be used such as the ones involved in DNNs mentioned in the previous section.
\e To summarize, the conclusion stated above and the two previous remarks are in favor of more investigation regarding \textit{conventional} approaches which is done in this paper.

\subsubsection{On the labeling problem in industrial context and the importance of feedback}\label{subsec_labelling} 
\noindent The presence of anomalies in equipment does not mean that these outliers can be detected in every single window of the time-series\footnote{Think about a default in the braking device which can be revealed only over those timestamps involving braking operations.}. In spite of this obvious fact, it is likely that the labels provided by the operator would refer to an \textit{anomalous} system as soon as the default is detected and until its handling by the maintenance operators, regardless of \textit{braking operations being undertaken or not}. This strongly biases the performance metrics of any model. \e 
Analogously, the early detection of anomalies negatively impacts the measured performance metrics of a model \cite{Dion2024}. Indeed, given the initially provided labels, an early detection increases the rate of \textit{False Positive} instances if the operators switch the label to \textit{anomalous} a bit lately, i.e only once the humanly measurable effects of the default are detected or because of the long periods separating two successive inspection rounds. \e The consequence of this is that very good models would be penalized for precisely being good in anticipating the human acknowledgment of the failure. Consequently, these labeling problems introduce high bias on the performance metrics when industrial-like time-series and labeling are involved, on one hand, and more importantly, they underline the importance of Operator Feedback Handling. 
\e The \textbf{Operator Feedback} is the mechanism by which a careful labeling (or re-labeling) is returned after an alarm is raised to inform an algorithm whether the sample under scrutiny is truly anomalous or not, or even when a non detected anomaly is discovered by the operators which has been considered by the anomaly detector as normal. 
\e 
The discussion above brings under light a property of an anomaly detector that is scarcely considered in the commonly provided comparisons while it is tremendously important to the adoption of any algorithm in real-life industry, namely the notion of \textbf{Feedback Friendliness}. A model is deemed feedback-friendly if it can be incrementally re-tuned to take into account an operator feedback regarding the diagnosis of a specific time-series window. \\ \\
Incorporating operator's feedback might be done using the so-called \textbf{Incremental Learning} \cite{Souiden2022, Zong2018, Bifet2009} (or batch learning) concept, where the model is regularly updated when a new batch of data is available. An approach to include these feedbacks into the proposed algorithm will be detailed below.

\section{Positioning of the proposed Anomaly Detector}\label{sec_pos_algo}
\noindent The guidelines that characterize the proposed anomaly detector can be summarized by the following requirements and items of concerns:
\begin{enumerate}[label=(\roman*)]
    \item We seek a design that exploits the \textbf{specificity of industrial time-series} that should obey some underlying, although unknown, physical laws. The latter introduce correlations between the \textit{state vector values} (represented by \textit{all the available sensors values at some instant together with their delayed versions}) and the possible \textbf{time increments of sensors values}. This is the reason why these increments play a crucial role in the forthcoming development.
    \item In order to fully implement the previous intuition, it is mandatory to come out with a \textbf{fully multivariate solution} (by opposition to those solutions that are mainly mono-sensors but which are extended to the multivariate context by simple features concatenation).
    \item We seek a design that is \textbf{robust to sensor incompleteness}. Indeed, with industrial time-series, in it not always the case that all the information that is needed to model the dynamics of the underlying process are included in the available data. This means that predicting the next values of the sensors is frequently not possible (if not irrelevant), or to say it differently, far too ambitious for the \textit{anomaly detection task}. That is why in the proposed solution, \textbf{the residual is not based on a reconstruction error of future signals}.
    \item Last but no least, in order to address common industrial expectations, a solution with a \textbf{high degree of explainability} should be preferred. Typical explanatory sentences might be: "\textit{there is something wrong with sensor i}" or "\textit{something unusual when the dynamics of sensor j is high}" or "\textit{sensor i goes beyond its normal domain when sensor j is going down slowly!}". Such additional explanatory hints might not precisely localize the issue but it can greatly help operators to generate working assumptions that can accelerate the diagnosis task.
\end{enumerate}
The concatenation of the previous four items enables a clear positioning of the proposed design w.r.t the existing categories of works recalled in Section \ref{subsec_sota}. More precisely: we believe that DNNs-based solutions are incompatible with items (iii) and (iv), and that Machine Learning solutions are not fully compatible with 1. Moreover the conclusion of \cite{Audibert2022} suggests that the non time-series particular nature of these approaches seems to penalize their performance in that specific context. \\ \\
As for conventional methods sub-categories, leaving aside the forecasting-based approaches (item (iii)), we believe that the decomposition methods are more oriented by the shape of the time-series rather than the link between the temporal increments and the state configuration. Two sub-categories remain that are tightly linked to our proposal which are the control-chart and the similarity search conventional approaches. The difference lies in the main role played by the sensors time increments in the construction of the associated \textit{control-charts}-like normality characterizations proposed in this contribution and the strong multivariate character involved by opposition to the similarity based approaches mentioned in Section \ref{subsec_conventional} which mainly concern cyclic mono-variate time-series. 

\section{Proposed algorithm}\label{sec_algo}
The key points in the proposed algorithm can be summarized as follows: 
\begin{enumerate}
    \item There is a mapping of the continuous values of the time-series toward \textit{ordered clusters} using a \textit{quantizer}. Given an integer $n_q \in \mathbb{N}\setminus\{0,1\}$, a $n_q$-quantizer is an application $\mathcal{Q} : \mathbb{R} \rightarrow \{0, \ldots, n_q - 1\}$ that associates an integer index to any real number. A typical way to \textit{fit} a quantizer using \textit{training data} is to match the values to the $k$-th $n_q$-quantile they belong to, with $k \in \{0, \ldots, n_q - 1\}$. \label{first_principle}
    \item Then, using a given time step $\Delta \in \mathbb{N}^*$, tuples are generated associating a quantization at a given time with the quantization $\Delta$ time units later. These tuples are called \textit{transition pairs} and represent images of the averagaed derivatives over the observation windows of length $\Delta$.\label{second_principle}
    \item For a given sensor and a visited transition pair, a \textit{configuration} vector is constructed, using both its past values and the present values of the other sensors, to serve as features for novelty detection. \label{third_principle}
\end{enumerate}
The following subsections propose a detailed description of the various steps of the algorithm ranging from the values quantization to the residuals generation. But let us first set some definitions and notation used in the sequel.
\subsection{Definitions and Notation}\label{subsec_notations}
\noindent Notice first of all that an exhaustive parameters dictionary is proposed in \ref{appendix_parameters} which provides a clear distinction between the algorithm's \textbf{hyper-parameters} that are provided by the users and the \textbf{model parameters} that results from the model's fitting.
\e Moreover, in the forthcoming development, the following notation is used: For any integer $N\in \mathbb{N}^*$, the notation $\overline{N}$ stands for the set of integer $\{1,\dots, N\}$. For a set of instants $\mathcal T$, the notation $s^{[i]}_\mathcal{T}$ stands for the values of sensor $i$ at the instants belonging to $\mathcal T$. $\tilde{s}$ stands for the scaled version of a time series $s$.
\e 
\noindent We assume that a \textit{training dataset} $\mathbb{D}_{train}$ is available, with $\mathcal{T}_{train}$ the set of timestamps present within, namely:
\begin{align}
\mathbb D_\text{train}:=& \Bigl\{s^{[i]}_{\mathcal{T}_{train}} \text{ \(|\) } i\in \overline{N_s}\Bigr\}\label{def_Dtrain} 
\end{align} 
\textbf{For a given sensor $i$}, these measurements are used to compute a $n_q$-quantizer $\mathcal{Q}^{[i]}$. First, the $i$-th sensor values are scaled to get $\tilde s^{[i]}_{\mathcal{T}_{train}}$. Then, with $v_1^{[i]}, \ldots, v_{n_q-1}^{[i]} \in \mathbb{R}$ being the ordered $n_q$-quantiles of the scaled $i$-th sensor training values, we define $n_q$ disjoint values intervals, namely:
\begin{equation}
    ]-\infty,v_0^{[i]}]\quad,\quad  ]v_0^{[i]}, v_1^{[i]}]\quad,\quad \\
     \ldots,\quad ]v_{n_q-1}^{[i]}, +\infty[ \label{def_interval} 
\end{equation}
which enables to define a $n_q$-quanting operator $ \mathcal{Q}^{[i]}$ s.t.:
\begin{align}
\begin{split}
    \mathcal{Q}^{[i]} := & \mathbb{R} \longrightarrow \{0, 1, \ldots, n_q-1\} \\
    & x \mapsto \begin{cases}
        0 & \text{ if } x \in ]-\infty, v_0^{[i]}] \\
        1 & \text{ if } x \in ]v_0^{[i]}, v_1^{[i]}] \\
        & \vdots \\
        n_{q}-2 & \text{ if } x \in ]v_{n_q-2}^{[i]}, v_{n_q-1}^{[i]}] \\
        n_q -1 & \text{ else}\\
    \end{cases}
\end{split}\label{def_quantizer}
\end{align}
Once such $n_q$-quantizer $\mathcal{Q}^{[i]}$ is defined for a sensor $i$, it is possible to create a quantized version $q^{[i]}_{\mathcal{T}_{train}}$ of the associated sensor time-series, namely:
\begin{equation}
    q^{[i]}_{\mathcal{T}_{train}} := \Bigl\{q^{[i]}_{t} = \mathcal{Q}^{[i]}(\tilde s^{[i]}_{t}) \text{ \(|\) } t \in \mathcal{T}_{train}\Bigr\}\label{def_quantization_cluster}
\end{equation}
Moreover, given a time step $\Delta \in \mathbb{N}^*$, the quantized value $\Delta$ timestamps ahead, $q^{[i]}_{t+\Delta}$ can be extracted. The couples of the form
\begin{equation}
    c^{[i]}_{t} := (q^{[i]}_{t}, q^{[i]}_{t+\Delta}) \in \{0, 1, \ldots, n_q-1\}^2 \label{def_transition}
\end{equation}
are referred to as \textit{transition pairs}. For a given sensor $i$, the set of the transition pairs over the training data is denoted by $c^{[i]}_{\mathcal{T}_{train}}$, namely
\begin{equation}
     c^{[i]}_{\mathcal{T}_{train}} := \Bigl\{c^{[i]}_{t} \text{ \(|\) } t \in \mathcal{T}_{train}\Bigr\} \label{def_transition_set}
     \end{equation}
\textbf{For a given sensor $i$}, a given time point $t \in \mathcal{T}_{train}$, and the associated transition pair $c^{[i]}_{t}$, a \textit{configuration vector} $w^{[i]}_{t}$ is defined by
\begin{equation}
w^{[i]}_{t} := \Bigl\{\tilde s^{[i]}_{t-\Delta+1}, \ldots, \tilde  s^{[i]}_{t-1}\Bigr\} \cup \Bigl\{\tilde  s^{[1]}_{t}, \ldots, \tilde s^{[N_s]}_{t}\Bigr\} \in \mathbb{R}^{\Delta + N_s - 1} \label{def_configuration_vector}
\end{equation}
which represents the concatenation of the delayed values of $\tilde s^{[i]}_t$ up to $\Delta$ previous samples and of the present values of all sensors. 
\begin{center}
\vskip -5mm
\begin{tikzpicture}
\node[rounded corners, fill=black!10, inner sep=4mm](P){
\begin{minipage}{0.45\textwidth}Notice that the definition of the configuration vector $w_t^{[i]}$ is \textbf{sensor-based} suggesting that a multi-variate sensor-based point-of-view will be adopted and this for each sensor. This is one of the major characteristics of the proposed algorithm 
\end{minipage} 
};
\end{tikzpicture}
\end{center}
For a given visited transition $c \in c^{[i]}_{\mathcal{T}_{train}}$, the following notation is used:
\begin{align}
    \mathcal{W}^{[i]}_c (\mathcal{T}_{train}) := &\Bigl\{ w^{[i]}_t \text{ \(|\) } t \in [\mathcal{T}_{train}]^{[i]}_{c} \Bigr\} \label{def_configuration_set}\\
    [\mathcal{T}_{train}]^{[i]}_{c} := & \Bigl\{ t \in \mathcal{T}_{train} \text{ \(|\) } c^{[i]}_{t} = c \Bigr\} \label{def_configuration_set_timestamps}
\end{align}
in order to denote the set of configurations $\mathcal{W}^{[i]}_c (\mathcal{T}_{train})$ for which the sensor $i$ witnesses some transition $c$ and its set of associated timestamps $[\mathcal{T}_{train}]^{[i]}_{c}$. 
\subsection{Normality characterizations over a training dataset}\label{subsec_normality}
\noindent Since anomaly detection is about detecting \textit{out-of-normality} configurations, let us define the three characterizations of normality that underlines the proposed anomaly detection algorithm. The normality characterization are computed based on three so called Normality Parameters (NP) that are first successively described before the residuals are precisely defined.
\subsubsection{\textsc{NP-1}: The set of transitions}\label{subsec_np1}
\noindent The first normality parameter viewed by the sensor $i$ is the set of transitions $c^{[i]}_{\mathcal{T}_{train}}$ itself. Indeed, a scenario involving a transition $c_{new}^{[i]} \notin c^{[i]}_{\mathcal{T}_{train}}$ at some time instant $t_{new}$ that has never been seen in the training data might be considered as anomalous. Gathering all the sets of transitions for all the sensors provides the first Normality Parameters that is defined by:
\begin{equation}
    \mathbf{NP}_1({\mathcal{T}_{train}}) := \Bigl\{c^{[i]}_{\mathcal{T}_{train}} \text{ \(|\) } i \in \overline{N_s} \Bigr\} \label{def_NP1}
\end{equation}

\subsubsection{\textsc{NP-2}: The transition-labeled bounds on excursions}\label{subsec_np2}
\noindent The second set of normality parameters viewed by the sensor $i$ is the bounds of the domains of variations conditioned by the transition values. More precisely, for a given sensor $i$ and a visited transition $c \in c^{[i]}_{\mathcal{T}_{train}}$, the following bounds are considered, with $w_j$ denoting the $j$-th component of a given configuration vector $w$\footnote{Note that, instead of the min and max operators, extreme values-robust ones could be used such as the percentiles 0.1\% and 99.9\%.} defined by \eqref{def_configuration_vector}:
\begin{align}
    \underline{b^{[i]}_c}(\mathcal{T}_{train}) := &\left[\underset{w \in \mathcal{W}^{[i]}_{c}(\mathcal{T}_{train})}{min} w_j\right]_{1 \leq j \leq \Delta + N_s} \in \mathbb{R}^{\Delta + N_s -1} \label{def_lower_bounds}\\
    \overline{b^{[i]}_c}(\mathcal{T}_{train}) := &\left[\underset{w \in \mathcal{W}^{[i]}_{c}(\mathcal{T}_{train})}{max} w_j\right]_{1 \leq j \leq \Delta + N_s} \in \mathbb{R}^{\Delta + N_s -1} \label{def_upper_bounds}
\end{align}
Gathering these bounds for all sensors provides the second Normality Parameter:
\begin{align}
    \mathbf{NP}_2({\mathcal{T}_{train}}) := &\Bigl\{\mathbf{NP}_2^{[i]}({\mathcal{T}_{train}}) \text{ \(|\) } i \in \overline{N_s}\Bigr\}  \quad \textrm{where} \label{def_NP2} \\
    \mathbf{NP}_2^{[i]}({\mathcal{T}_{train}}) := &\Bigl\{(\underline{b^{[i]}_c}(\mathcal{T}_{train}), \overline{b^{[i]}_c}(\mathcal{T}_{train})) \text{ \(|\) } c \in c^{[i]}_{\mathcal{T}_{train}} \Bigr\} \label{def_NP2_ith}
\end{align}
\subsubsection{\textsc{NP-3}: The transition-labeled sensors configurations}\label{subsec_np3}
\noindent The third set of normality parameters viewed by the sensor $i$ is the set of uncorrelated configuration vectors conditioned by the transition values. More precisely, for a given sensor $i$ and a visited transition $c \in c^{[i]}_{\mathcal{T}_{train}}$, a smallest  set $\widetilde{\mathcal{W}}^{[i]}_{c}(\mathcal{T}_{train})$ of configurations is generated by removing the highly correlated configuration vectors:
\begin{equation}
\centering
\begin{aligned}
    \widetilde{\mathcal{W}}^{[i]}_{c}(\mathcal{T}_{train}) := & \Bigl\{w \in \mathcal{W}^{[i]}_{c}(\mathcal{T}_{train}) \text{ \(|\) } \\
    & \max(|\textbf{corr}(w, \mathcal{W}^{[i]}_{c}(\mathcal{T}_{train}))|) < \eta \Bigr\} \quad\textrm{where} \label{def_configuration_set_lowcorr}
\end{aligned}
\end{equation}
\begin{equation}
\centering
\begin{aligned}
    \textbf{corr}(w, \mathcal{W}^{[i]}_{\mathcal{T}_{train}}) := & \Bigl\{\textbf{corr}(w, w') \text{ \(|\) } \\
    & w' \in \mathcal{W}^{[i]}_{c}(\mathcal{T}_{train}), w \neq w' \Bigr\} \label{def_correlation}
\end{aligned}
\end{equation}
Gathering these sets of uncorrelated configurations for all sensors, we can compute the last Normality Parameter:
\begin{align}
    \mathbf{NP}_3({\mathcal{T}_{train}}) := &\Bigl\{\mathbf{NP}_3^{[i]}({\mathcal{T}_{train}}) \text{ \(|\) } i \in \overline{N_s} \Bigr\}  \quad \textrm{where}  \label{def_NP3} \\
    \mathbf{NP}_3^{[i]}({\mathcal{T}_{train}}) := &\Bigl\{\widetilde{\mathcal{W}}^{[i]}_{c}(\mathcal{T}_{train}) \text{ \(|\) } c \in c^{[i]}_{\mathcal{T}_{train}} \Bigr\} \label{def_NP3_ith}
\end{align}
The cardinality of this set of normality parameters cannot be determined a priori. However, it is shown in Section \ref{sec_examples} that this cardinality can generally be quite low. This suggests that the underlying physical
laws combined with the transition-induced clusters reduce the number of really different (uncorrelated) sensors' configurations for a given transition. Nonetheless, it can be necessary to control the cardinality of these sets. To achieve this, a mechanism is discussed in \ref{appKmeans}. 
\e 
The computation of the three Normality Parameters discussed above can be summarized by the following Algorithm \ref{alg1page}.

\begin{algorithm}
\caption{Fit the model}\label{alg1page}
\begin{algorithmic}[1]
\Require $\mathcal{T}_{train} \neq \emptyset$, $n_q \geq 2$, $\Delta \geq 1$, $\eta \in ]0, 1[$ 
\State Scale the values associated to $\mathcal{T}_{train}$
\For{$i \in  \overline{N_s}$}
\State Compute the $n_q$-quantizers $\mathcal Q^{[i]}$ 
\State Compute the transitions $c_{\mathcal{T}_{train}}^{[i]}$ using $\Delta$ [$=\mathbf{NP_1^{[i]}}$]
\For{$c\in c_{\mathcal{T}_{train}}^{[i]}$}
\State Compute $\mathcal{W}^{[i]}_{c}(\mathcal{T}_{train})$ 
\State Compute $\underline{b^{[i]}_c}(\mathcal{T}_{train})$ and $\overline{b^{[i]}_c}(\mathcal{T}_{train})$ [$\in \mathbf{NP_2^{[i]}}$]
\State Compute $\widetilde{\mathcal{W}}^{[i]}_{c}(\mathcal{T}_{train})$  using $\eta$ [$\in \mathbf{NP_3^{[i]}}$]
\EndFor
\EndFor
\end{algorithmic}
\end{algorithm}
\noindent As there is no trivial way to estimate the perfect hyper-parameters from a blind approach to industrial data, a study on the hyper-parameters selection is discussed in \ref{subsec_hyperpameters_selection}, along with the construction of a \textit{rule-of-thumb}. \\ \\
In order to respect a \textit{blind} approach as an evaluation guideline for the benchmark proposed in Section \ref{sec_bench}, these rules of thumb are not used.
\subsection{Incremental Learning potential}\label{subsec_il}
\noindent Monitoring industrial equipment health is often associated with the notion of \textbf{operator feedback} (see Section \ref{subsec_labelling} for a detailed discussion). These feedbacks generally amount at re-labeling of specific periods of the time-series, which can alter the perceived performance of an already-fitted algorithm. \\ \\ 
As re-training the algorithm from scratch using the new labels can be time-prohibitive or even infeasible for an in-use industrial data stream monitoring, it has been strengthened at the end of Section \ref{subsec_labelling} that this can be handled via an \textbf{Incremental Learning} step. As far as the proposed algorithm is concerned, the incremental learning step amounts at modifying the previously presented Normality Parameters by either increasing or decreasing the corresponding sets. This is detailed in the following sub-sections.
\subsubsection{Incremental Learning for normality-stamped feedback}\label{subsec_il_increase_norm}
\noindent This concerns new data that is labeled as normal by the \textit{operator feedback}. This might be about the integration of a \textbf{new context of use} or of a \textbf{drift in normality} \cite{Gama2004} to the initial model. The incremental learning in this case amounts at \textbf{increasing} the model Normality Parameters sets. \\ \\
Informally, such an update can be described as follows: Given an initial model fitted on some training dataset $\mathcal{T}_{train}$ and a new sub-time-frame $\mathcal{T}_{new}$ that is \textbf{labeled as normal} by the operator feedback, the Normality Parameters are updated as follows:
\begin{itemize}
    \item $\mathbf{NP}_1$: The transitions in $\mathcal{T}_{new}$ are concatenated to the already learned ones in $\mathcal{T}_{train}$.
    \item $\mathbf{NP}_2$: For the common transitions between $\mathcal{T}_{train}$ and $\mathcal{T}_{new}$, the bounds are updated. Otherwise, new bounds are added referenced by the new transitions.
    \item $\mathbf{NP}_3$: For the common transitions between $\mathcal{T}_{train}$ and $\mathcal{T}_{new}$, the vectors ensembles are updated. Otherwise, new bounds are added referenced by the new transitions.
\end{itemize}
A more rigorous definition of this step can be found in \ref{il_increase_norm}.

\subsubsection{Incremental Learning for anomaly-stamped feedback} \label{subsec_il_reduce_norm}
\noindent Alternatively, the \textbf{operator feedback} can be about an \textbf{undetected anomaly}. This case, that is not usually covered in literature induces to the challenging task of reducing the normality space without increasing the risk of False Positive rate.  \\ \\
Informally, such an update can be described as follows: Given an initial model fitted on some training dataset $\mathcal{T}_{train}$, a new sub-time-frame $\mathcal{T}_{new}$ concerned by the operator and a pre-existing threshold  $\zeta$ used in the correlation criterion:
\begin{itemize}
    \item $\mathbf{NP}_3$: For the common transitions between $\mathcal{T}_{train}$ and $\mathcal{T}_{new}$, the $\mathcal{T}_{train}$ highly correlated vector (correlation greater than $\zeta$) to the $\mathcal{T}_{new}$ configuration vectors are \textbf{forgotten}. If the result of the process leaved a set of configuration vectors for a given transition empty, then its affiliated transition is \textbf{dropped}. Notice that this process leaves unchanged all the $\mathcal{T}_{train}$ that are not common with $\mathcal{T}_{new}$, for those transitions that are only present in $\mathcal{T}_{new}$, they are obviously not memorized.
    \item $\mathbf{NP}_2$: For the \textbf{remaining} common transitions between $\mathcal{T}_{train}$ and $\mathcal{T}_{new}$, the bounds are updated. There is no change on the $\mathcal{T}_{train}$-only ones.
    \item $\mathbf{NP}_1$: The list of transitions is reduced to the $\mathcal{T}_{train}$ remaining ones.
\end{itemize}
A more rigorous definition of this step can be found in \ref{il_reduce_norm}.
\e Now that the Normality Parameters are defined and their updating mechanism upon arrival of operator feedback is described. It is time to define how the residuals serving in the overall assessment of normality are defined. This is the aim of the next section. 
\subsection{Computing the residuals}\label{subsec_residuals}
\noindent Associated to the normality parameters defined in Sections \ref{subsec_np1}, \ref{subsec_np2} and \ref{subsec_np3}, three corresponding residuals are defined for a better explainability of the anomalous time-series. The residuals are computed for time-series that last over a time window including $n_{pred}$ successive measurements. Note that in order for the underlying time window to contain at least one transition, the following constraint should be respected:
\begin{equation}
    n_{pred} > \Delta \label{def_constraint_win_size}
\end{equation}
Notice that for any valid choice of $n_{pred}$ it is possible to compute the associated residuals by using the same normality parameters introduced in Section \ref{subsec_normality}. The window length plays the role of \textit{aggregating parameter} that can be adapted to the presumed time scale of the faults visibility and persistence. \\ \\
The three residuals are respectively denoted by $r_{\textrm{trans}}$ (transitions), $r_{\textrm{bound}}$ (bounds) and $r_{\textrm{conf}}$ (configurations) since they are expected to reveal three different kinds of anomalies consequences. As the \textbf{faults explainability} is one of the key strength of this algorithm, \textbf{the residuals will be computed for each sensor}, without aggregation. Note that even if the residuals are displayed for a specific sensor, the computing steps are based on a \textbf{multi-sensor view}.\\ \\
\textit{All the computations reported in the sequel are focused on a given $n_{pred}$ and should be repeated over a sliding window to account for an entire dataset}.

\subsubsection{Anomalous transitions-related residual $r_{\textrm{trans}}$}\label{subsec_rtrans}
\noindent Denoting by $\mathcal{T}_{pred}$ the time interval over which the residuals are to be predicted, and consisting of $n_{pred}$ contiguous sampling instants, the residual $r_{\textrm{trans}}^{[i]}$ associated to anomalous transitions for a sensor $i$ is defined as follows:
\begin{equation}
    r_{\textrm{trans}}^{[i]}(\mathcal{T}_{pred}) := \frac{1}{n_{pred}} \Bigl|\Bigl\{t \in \mathcal{T}_{pred} \text{ \(|\) } c^{[i]}_t \notin c^{[i]}_{\mathcal{T}_{train}} \Bigr\} \Bigr| \label{def_res_trans}
\end{equation}
where $c^{[i]}_{\mathcal{T}_{train}}$ is the set of \textit{normal} transitions computed using the training data and retained in the sensor $i$ first normality parameters set $\mathbf{NP}_1(\mathcal{T}_{train})^{[i]}$. More clearly, the residual $r_{\textrm{trans}}$ corresponds to the normalized\footnote{with regard to the length $n_{pred}$ of the prediction interval $\mathcal{T}_{pred}$.} number of \textbf{novel transitions} (observed at instants $t \in \mathcal{T}_{pred}$ belonging to the prediction time window). Notice that novel transitions can be raised following a context of use that involves transitions that were not present in the training data, but it can be also true that these unexpected transitions are the result of faults impacting the health of the equipment.
\subsubsection{Anomalous domain-related residuals $r_{\textrm{bound}}$}\label{subsec_rbound}
\noindent Contrary to $r_{\textrm{trans}}$, the residual $r_{\textrm{bound}}$ corresponds to \textit{normal} transitions $c$ for a sensor $i$ for which anomalous values of some sensors\footnote{Viewed by sensor $i$: this includes all the original sensors to which the delayed versions of sensor $i$ are added.} in the configuration vectors sets are observed. Namely, these anomalous values do not belong to the domains defined by the normality parameters $\overline{b^{[i]}_c}$ and $\underline{b^{[i]}_c}$ for that specific transition $c \in c^{[i]}_{\mathcal{T}_{train}}$ (as retained in $\mathbf{NP}_2(\mathcal{T}_{train})^{[i]}$). Consequently, this residual is defined, for a sensor $i$, by:
\begin{align}
    &r_{\textrm{bound}}^{[i]}(\mathcal{T}_{pred}) :=  \frac{1}{n_{pred}} \sum_{t \in \mathcal{T}_{pred}} \delta^{[i]}(t) \quad \textrm{where} \label{def_res_bound} \\
    \begin{split}
    &\delta^{[i]}(t) :=  \frac{1}{N_s + \Delta - 1} \mathbf{dist}\Bigl(w^{[i]}_{t}, [\underline{b}^{[i]}_{c^{[i]}_{t}}(\mathcal{T}_{train}), \overline{b}^{[i]}_{c^{[i]}_{t}}(\mathcal{T}_{train})] \Bigr) \label{def_bound_delta}
\end{split}
\end{align}
where $w^{[i]}_t$ is the configuration vector at instant $t \in \mathcal{T}_{pred}$ and $\underline{b}^{[i]}_{c^{[i]}_{t}}, \overline{b}^{[i]}_{c^{[i]}_{t}}$ the $\mathbf{NP}_2^{[i]}(\mathcal{T}_{train})$-bounds corresponding to the transition $c \in c^{[i]}_{\mathcal{T}_{train}}$ observed at the same instant. \\
The distance mentioned is the one between a vector $w^{[i]}_t \in \mathbb{R}^{\Delta + N_s-1}$ and the normal hypercube in $\mathbb{R}^{\Delta + N_s-1}$, which is defined by the second argument $[\underline{b}^{[i]}_{c^{[i]}_{t}}, \overline{b}^{[i]}_{c^{[i]}_{t}}]$. This distance can be derived in a straightforward manner from any scalar distance between a real $\xi$ and a scalar interval $[\underline{\xi}, \overline{\xi}]$ such as the one defined by:
\begin{equation}
    \mathbf{dist}\Bigl( \xi, [\underline{\xi}, \overline{\xi}] \Bigr) :=  \frac{\max\{0, \underline{\xi} - \xi \} + \max\{0, \xi - \overline{\xi}\}}{\overline{\xi} - \underline{\xi} + \epsilon}   \label{def_distance}
\end{equation}
When this residual takes non-zero values, it underlines that during some normal transition of sensor $i$, at least a sensor (say $j$) took values that are outside the normal domain observed in the training dataset. This can be an indicator of some changes in the transfer (gain) coefficient involving the way the sensor $j$ appears in the r.h.s. of the unknown physical relationship that governs the increments of sensor $i$.
\subsubsection{Anomalous configuration-related residual $r_{\textrm{conf}}$}\label{subsec_rconf}
\noindent The last residual, namely $r_{\textrm{conf}}$, is oriented towards the detection of changes in the relative configurations of the different sensors (including the additional delay related ones) for a visited set of transitions affecting the different original sensors. As explained in Section \ref{subsec_np3}, for a given sensor $i$ experimenting a transition $c^{[i]}_t \in c^{[i]}_{\mathcal{T}_{train}}$ at some instant $t \in \mathcal{T}_{pred}$, the \textit{normality} means that there is a vector in $\widetilde{\mathcal{W}}^{[i]}_{c^{[i]}_t}(\mathcal{T}_{train})$ which is highly correlated to the current measurement configuration vector $w^{[i]}_t$. This suggests the following definition of the residual, for a sensor $i$:
\begin{align}
  r_{\textrm{conf}}^{[i]}(\mathcal{T}_{pred}) &:=  1 - \min_{c \in c^{[i]}_{\mathcal{T}_{pred}}} \left( \frac{1}{\Bigl|\mathcal{W}^{[i]}_{\mathcal{T}_{pred}}\Bigr|} \right.\label{def_res_conf}
  \\ & \left.  \sum_{w \in \mathcal{W}^{[i]}_{\mathcal{T}_{pred}}} \max\left(\Bigl|\mathbf{corr}(w,  \widetilde{\mathcal{W}}^{[i]}_{c}(\mathcal{T}_{train})\Bigr|\right) \vphantom{\frac{1}{\Bigl|\mathcal{W}^{[i]}_{\mathcal{T}_{pred}}\Bigr|}} \right) \nonumber
\end{align}
where $c^{[i]}_{\mathcal{T}_{pred}}$ represents all the visited transitions for the sensor $i$ during $\mathcal{T}_{pred}$, $\mathcal{W}^{[i]}_{\mathcal{T}_{pred}}$ is the set of configuration vectors for a specific transition $c$ amongst them, $\widetilde{\mathcal{W}}^{[i]}_{c}(\mathcal{T}_{train})$ is the set of the uncorrelated normal configuration vectors for the same specific transition $c$ (retained in the $\mathbf{NP}_3(\mathcal{T}_{train})^{[i]}$) and the correlation mentioned is the one defined in (\ref{def_correlation}).\\ \\
Obviously this residual is expected to reveal a change in the relative configuration of sensors inducing the same increment of the sensor $i$. Since a summation on $i$ is used, the residual accounts for any change of the unknown underlying relationship governing all the involved dynamics.

\section{Illustrative examples}\label{sec_examples}
\noindent In this section, the behavior of the residuals generated by the above described algorithm is first examined on an illustrative example for a specific setting of the hyper-parameters before being analyzed through an extended statistical comparison with some other algorithms in Section \ref{sec_bench}.

\subsection{Example 1: Detecting parameters changes in Lorentz attractor}\label{subsec_lorentz}
\noindent In this first example\footnote{The datasets used in this example are accessible at \url{https://github.com/mazenalamir/lorentz-atractor-blind-fault-detection-dataset}}, the ability of the proposed framework to detect slight changes in the parameters of a dynamical system is investigated. To do so, the following well-known Lorentz model is used:
\begin{subequations}\label{lorentz}
\begin{align}
\dot x_1&={\color{orange} \sigma}(x_2-x_1) \label{eq1lorentzesigma}\\
\dot x_2&=x_1({\color{red} \rho}-x_3)-x_2 \label{eq2lorentzrho}\\
\dot x_3&=x_1x_2-{\color{blue} \beta} x_3 \label{eq3lorentzbeta}
\end{align} 	
\end{subequations}
where \textcolor{orange}{$\sigma$}, \textcolor{red}{$\rho$} and \textcolor{blue}{$\beta$} are the parameters that might be affected by unpredictable changes one would like to detect thanks to the anomaly detector, using the two measurements sensors defined by $s_1 = x_1$ and $s_2 = x_3$, namely:
\begin{equation}
    s := [x_1, x_3]
\end{equation}
\begin{figure}
    \centering
    \includegraphics[scale=0.25]{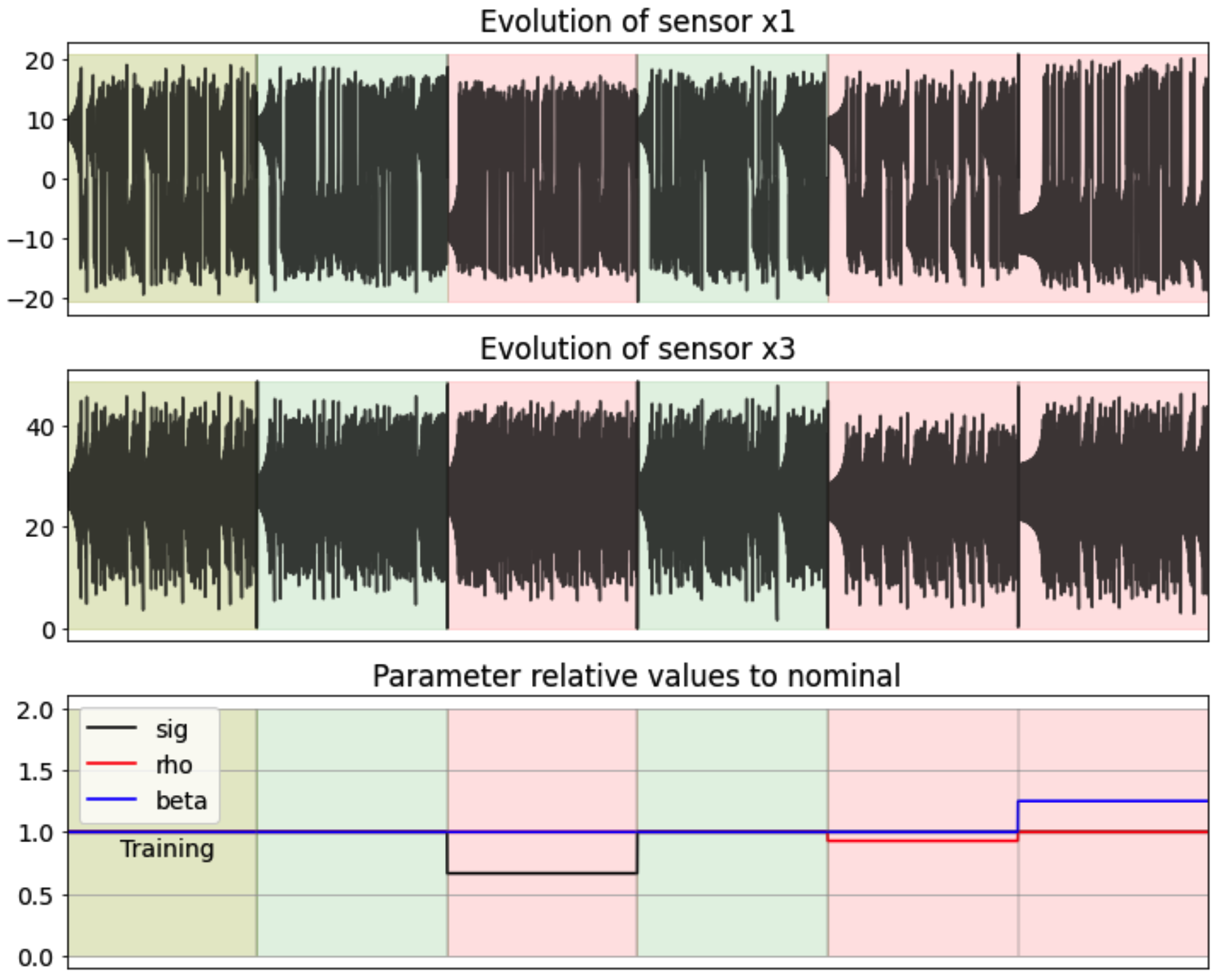}
    \caption{\textbf{Example 1:} Time-series representing the two sensors evolution and the associated values of the system's parameters. The green intervals represent normal behavior, while red ones represent faulty behavior. The bottom plot displays the parameters variations from the nominal values used for normal simulations.}\label{fig_lorentz_overview}
\end{figure}
The nominal values of the parameters are given by {$\sigma = 12$}, {$\rho = 28$} and {$\beta = \frac{8}{3}$}. The working time-series are depicted on Figure \ref{fig_lorentz_overview} where six different intervals of time-series are concatenated, each of which generated using different initial conditions and different values of the parameters. \\ \\
More precisely, the \textbf{green intervals} on Figure \ref{fig_lorentz_overview} correspond to nominal values of the parameters begin used while different initial states are used, leading to obviously different time-series because of the very nature of the oscillator which is not open-loop asymptotically stable. These intervals correspond to a \textbf{normal behavior}. \\ \\
The \textbf{red regions} correspond to different de-tuned sets of parameters as shown in the last row of plots, where the evolution of the three parameters displayed after normalizing by the nominal values mentioned above (hence 1 corresponds to the nominal values). To be more precise, in the first anomalous region (third interval), \textcolor{orange}{$\sigma$} takes a value that is 30\% lower than the nominal value (8 instead of 12). In the second anomalous interval (the fifth interval), the \textcolor{red}{$\rho$} parameter is 7\% lower than the nominal value (26 instead of 28). Finally, the last interval shows \textcolor{blue}{$\beta$} taking a value 25\% above the nominal one ($10/3$ instead of $8/3$). As mentioned, these intervals represent multiple \textbf{faulty behaviors}.\\ \\
Note the these relatively small changes are chosen in order to get raw time-series that cannot easily be distinguishable as it can be observed in Figure \ref{fig_lorentz_overview}. \\ \\
The time-series are obtained by simulating (\ref{lorentz}) using 4th-order Runge-Kutta method with a sampling period of $10^{-2}$. As for the anomaly detector setting, the following set of hyper parameters are used, accordingly to a rule of thumb that are discussed in \ref{subsec_hyperpameters_selection}:
\begin{equation}
n_q=20,\quad  \Delta=20, \quad \eta=0.95 \quad \epsilon = 1 \label{lorentzparam1}
\end{equation} 
\subsubsection{Results}
\begin{figure*}[ht]
\centering
\begin{minipage}{.45\textwidth}
  \centering
  \includegraphics[width=0.85\textwidth]{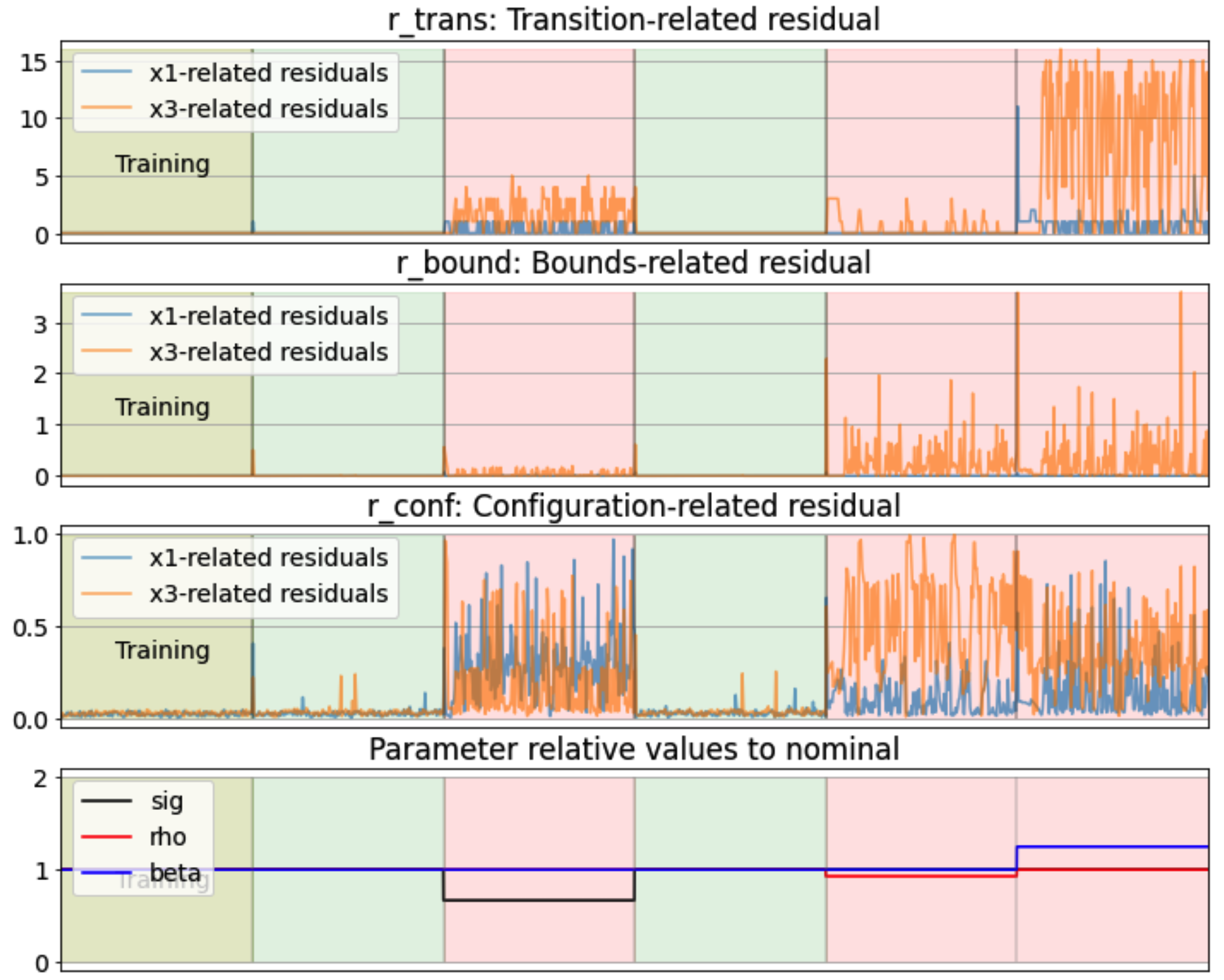} 
  \caption{\textbf{Example 1:} Evolution of the family of residuals viewed by sensor $x_1$ and $x_3$ when both sensors are available for the construction of the anomaly detector associated to the hyper-parameters given by \eqref{lorentzparam1}.}\label{fig_lorentz_results} 
\end{minipage}%
\hspace{0.05\textwidth}
\begin{minipage}{.45\textwidth}
  \centering
  \includegraphics[width=0.85\textwidth]{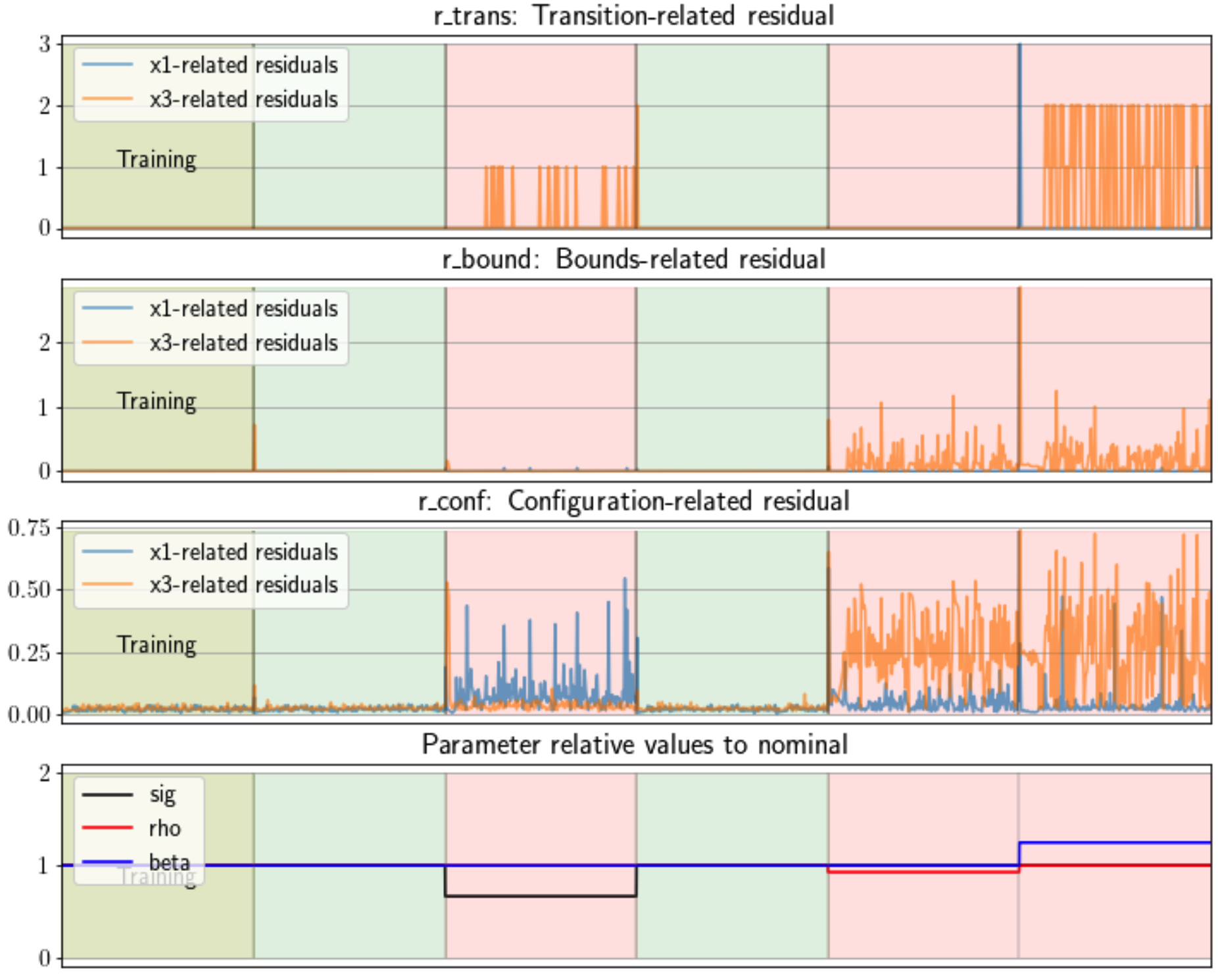} 
  \caption{\textbf{Example 1}: Evolution of the family of residuals viewed by sensor $x_1$ and $x_3$ when both sensors are available for the construction of the anomaly detector associated to the hyper-parameters given by \eqref{lorentzparam1} \textbf{except that $\mathbf{n_q=8}$ is used instead of $\mathbf{20}$}.}\label{fig_lorentz_results_nqlower} 	
\end{minipage}
\end{figure*}

\begin{figure}
    \centering
    \includegraphics[scale=0.38]{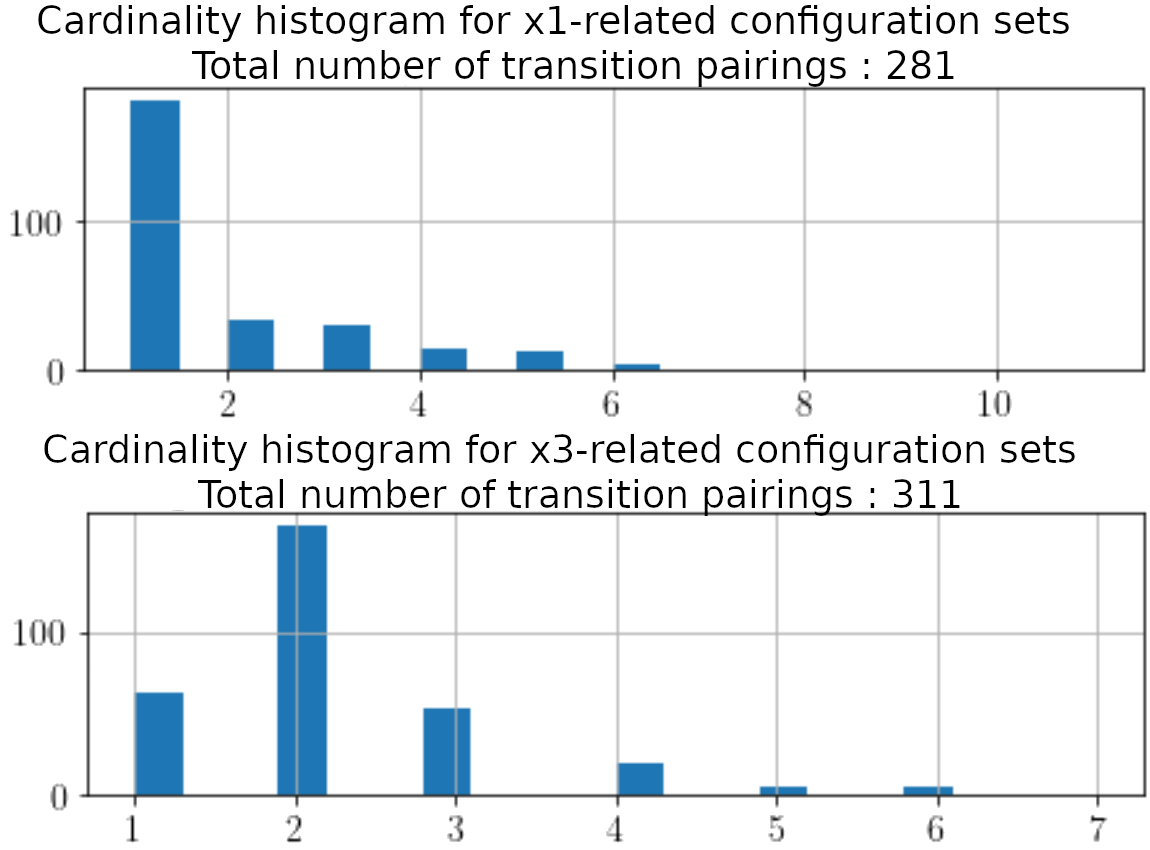}
    \caption{\textbf{Example 1:} Histograms of the cardinalities of the fitted sets of configuration $\mathcal{W}^{[x_1]}$ and $\mathcal{W}^{[x_3]}$ among the 281 transitions of quantized $x_1$ and 311 transitions of the quantized $x_3$. The anomaly detector's hyper-parameters are given by (\ref{lorentzparam1}).}\label{fig_lorentz_card_nq20}
\end{figure}

\begin{figure}
    \centering
    \includegraphics[scale=0.38]{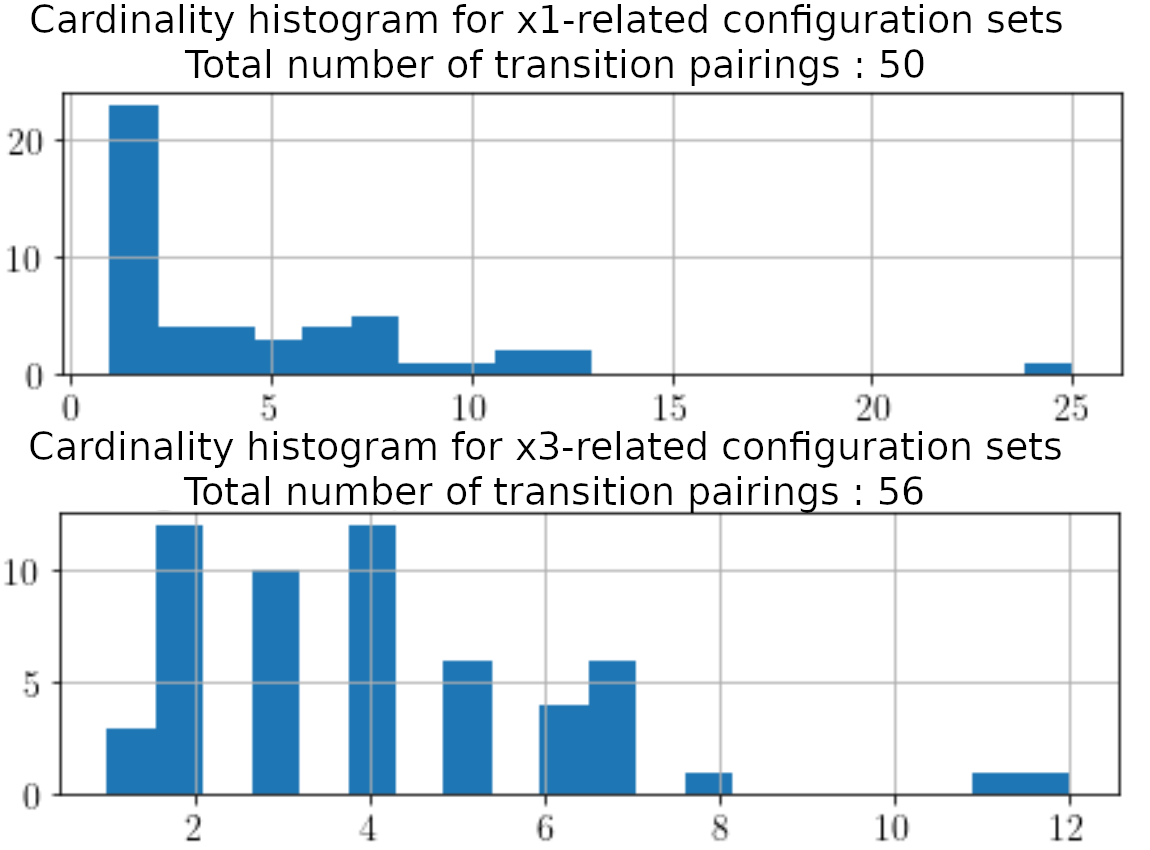}
    \caption{\textbf{Example 1:} Histograms of the cardinalities of the fitted sets of configuration $\mathcal{W}^{[x_1]}$ and $\mathcal{W}^{[x_3]}$ among the 50 transitions of quantized $x_1$ and 56 transitions of the quantized $x_3$. The anomaly detector's hyper-parameters are given by (\ref{lorentzparam1}) \textbf{except that $\mathbf{n_q=8}$ is used instead of $\mathbf{n_q = 20}$}.}\label{fig_lorentz_card_nq8}
\end{figure}

\noindent The residuals delivered by the fitted anomaly detector using a window of length $n_{pred} = 100$ are shown in Figure \ref{fig_lorentz_results} which clearly shows that over each of the anomalous regions, at least one of the residual is clearly above the level it takes over the normal regions. \\ \\
Moreover, in order to appreciate the impact of the choice of the quantization parameter $n_q$, the same results as the one shown in Figure \ref{fig_lorentz_results} are given in the case where $n_q = 8$ is used instead of $n_q = 20$. This leads to the results displayed in Figure \ref{fig_lorentz_results_nqlower}, on which we can notice  that the detection is still possible although $r_{\textrm{trans}}$ increase is far less important, suggesting a possible sensitivity of the success to random factors. But on the other hand, the detection of changes in \textcolor{orange}{$\sigma$} is now more clearly associated to the transition on $x_1$ as it is obvious from (\ref{eq1lorentzesigma}). On the other hand, the residual of the green normal regions shows also less burst, leading to less potential false positive diagnosis. This discussion clearly underlines that $n_q$ is a hyper-parameter to be finely tuned in the design process. \\
\begin{rmk}[on the explainability]\ \\
An interesting feature that can be observed is the high sensitivity of the transition-related residual $r_{\textrm{trans}}$ viewed by the sensor $x_3$ when the parameter \textcolor{blue}{$\beta$} is anomalous. This simply reflects (\ref{eq3lorentzbeta}) which clearly involves a direct impact of \textcolor{blue}{$\beta$} on the transitions of sensor $x_3$. It is also noticeable that for this example, and as long as the associated relatively small changes introduces on the system parameters are concerned, the bound-related residual $r_{\textrm{bound}}$ \textit{viewed by the sensor $x_1$} does not reveal any anomalies while the configuration-related residuals $r_{\textrm{conf}}$ \textit{viewed by both sensors} are sensitive to all anomalies.  \hfill $\spadesuit$
\end{rmk}
\noindent Figure \ref{fig_lorentz_card_nq20} shows the statistics of the cardinalities of the sets of configuration $\mathcal{W}^{[x_1]}$ and $\mathcal{W}^{[x_3]}$ among the 281 transitions of quantized $x_1$ and 311 transitions of the quantized $x_3$. Notice that the majority of these cases corresponds to a single vectors and that the cardinality rarely goes beyond 6, while the elements of these sets lie in $\mathbb{R}^{21}$ (since $\Delta + N_s - 1 = 20 + 2 - 1 = 21$).Notice that these cardinalities are intimately linked to the correlation threshold being used, namely $\eta = 0.95$. \\ 
\\
\noindent Finally, in order to appreciate the impact of the choice of the quantization parameter $n_q$, the same results as the one shown in Figure \ref{fig_lorentz_card_nq20} are given in the case where $n_q = 8$ is used instead of $n_q = 20$. This leads to the results displayed in Figure \ref{fig_lorentz_card_nq8}, which displays the obvious fact that when less quantized levels are used much less transition pairings occurs, but that inside each transition cluster the number of uncorrelated configuration vectors increases, leading to the cardinality of their sets of representatives sets to also increase. 

\begin{rmk}[on the residuals complementarity]\ \\
Notice that on Figure \ref{fig_lorentz_card_nq8} every faulty intervals cannot be derived from the observation of a single residual. Indeed, working with the sensor $x_3$ (in blue), the residual $r_{\textrm{conf}}$ is excited for the \textcolor{red}{$\rho$} and \textcolor{blue}{$\beta$} anomalies but is blind to the \textcolor{orange}{$\sigma$} one, whereas the residual $r_{\textrm{trans}}$ is excited for the \textcolor{orange}{$\sigma$} anomaly. Moreover, combining both the $r_{\textrm{trans}}$  and $r_{\textrm{conf}}$ for the sensor $x_3$ allows to discriminate between a \textcolor{red}{$\rho$} anomaly and a \textcolor{blue}{$\beta$} anomaly. \hfill $\spadesuit$ \label{rmk_complementarity}
\end{rmk}
\subsubsection{Illustration of the Incremental Learning process} \label{ex_il_increase_norm}
\noindent In order to illustrate the efficiency of the incremental learning process described in Section \ref{subsec_labelling} and the implementation proposed in Section \ref{subsec_il}, let us assume that an \textbf{operator} provided the following \textbf{feedback}:
\begin{center}
\begin{tikzpicture}
\node[rounded corners, fill=black!10, inner sep=4mm](P){
\begin{minipage}{0.45\textwidth}The first red region in Figure \ref{fig_lorentz_results}, which is associated to a non-nominal value of \textcolor{orange}{$\sigma$}, is in fact not a faulty behavior but rather a \textbf{normal context of use} that was not present in the training data.
\end{minipage} 
};
\end{tikzpicture}
\end{center} 

\begin{figure}
    \centering
    \includegraphics[scale=0.35]{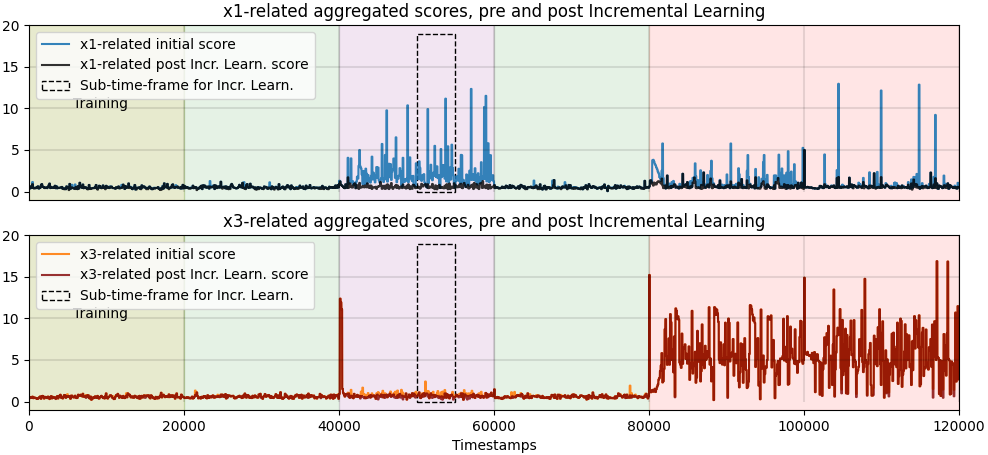}
    \caption{\textbf{Example 1:} Comparison of the algorithm output before and after an Incremental Learning step on the boxed sub-time-frame. For the sake of clarity, each sensor scores have been aggregated using the residuals maximum value for each timestamp after a normalization against the training part. The mislabeled period have been highlighted in purple. Notice that the burst that appears systematically at the boundaries of the domains are due to the discontinuity induced by the transition between the concatenated experiments that are not representative. }\label{fig_qboard_il_increase}
\end{figure}
\noindent The aggregated residuals after an Incremental Learning step are shown in Figure \ref{fig_qboard_il_increase}, with a detail sensor by sensor and the same hyper-parameters as \ref{lorentzparam1}. Notice how the $x_1$-related output is almost flat for the mislabeled period, whereas it was the main informant before the Incremental Learning step. Overall, while the performance of the $x_1$-related residuals downgrade slightly on the true anomalous areas, the algorithm proposes a precise discrimination between operator-informed true normal and true anomalous periods. 

\subsection{Example 2: Automotive Electronic Throttle Control} \label{subsec_etc}
\begin{figure}
  \centering
  \includegraphics[scale=0.25]{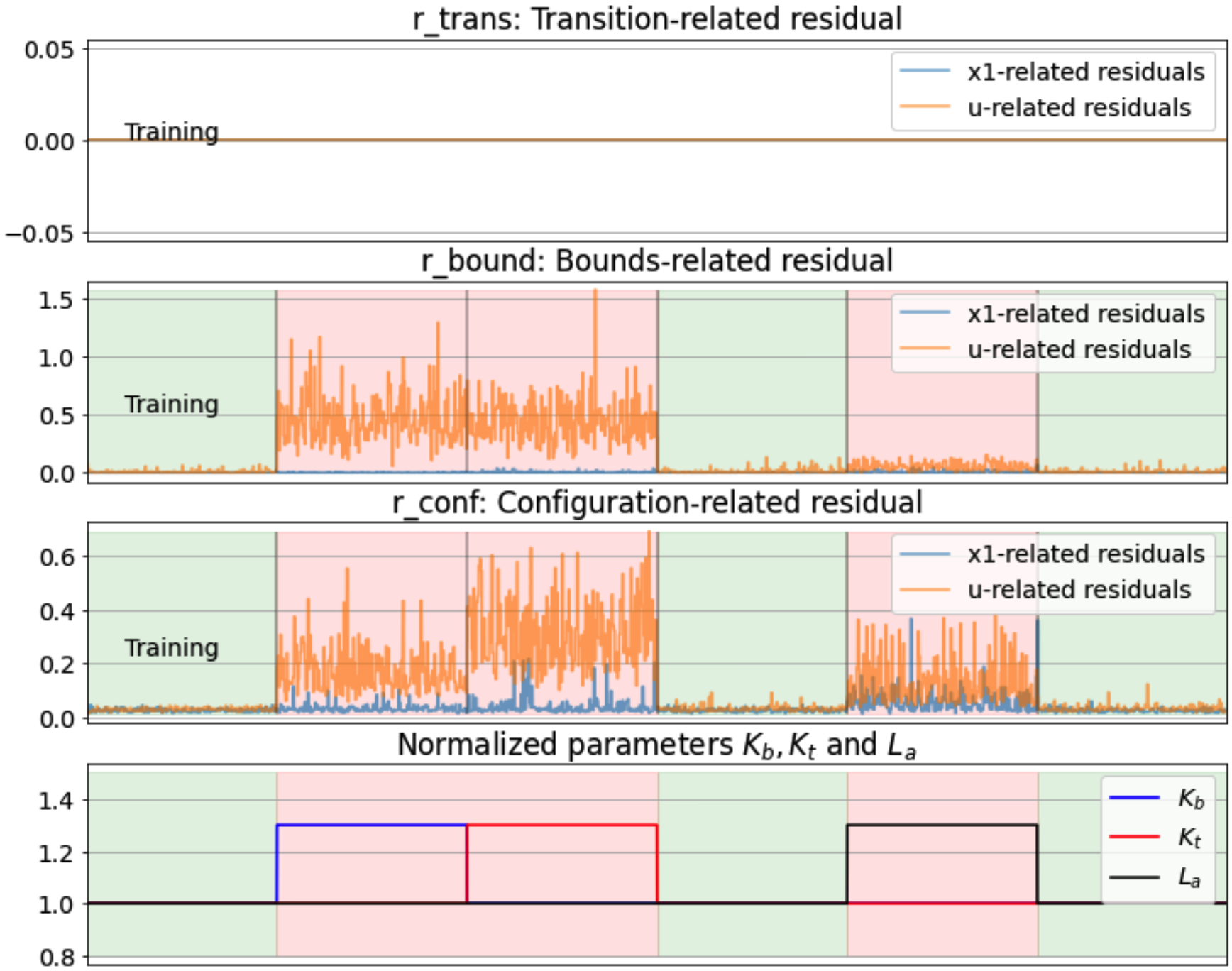} 
  \caption{\textbf{Example 2:} Evolution of the family of residuals viewed by sensor $\theta=x_1$ and $i_a=u$ when both sensors are available for the construction of the anomaly detector associated to the hyper-parameters given by \eqref{etcparam}.}\label{fig:etc_results_1}
\end{figure}

\noindent In this second example\footnote{The datasets of this example are freely accessible at \url{https://github.com/mazenalamir/ETC_dataset_BFD}}, the very common situation is investigated, where a controlled system is involved which is based on the set of nominal parameters of the system. In this case, a very closed-loop nature implies that the \textbf{feedback control} could hide some of the consequences of the parametric changes representing anomalies. Moreover, the variability of contexts materializes in the values of the set-point changes that can be applied and fed to the \textbf{control feedback algorithm}\footnote{Note that the word \textit{feedback} here does not have the same meaning as the one used in Section \ref{subsec_labelling}, where it is referred to the maintenance operator feedback. Here, feedback is used in the context of control systems to designate the action by which the controller reacts to the measurements in order to force the system to follow some desired behavior.}. \e 
A schematic view of the system is shown in Figure \ref{fig_etc} where the angle $\theta$ that controls the airflow is controlled via a DC servo-motor controlled by the applied armature voltage $e_a$, which is itself governed by the armature current $i_a$. Interest readers can refer to \cite{Conatser2004} for more technical details although a sketch of the underlying equations is given below for the sake of completeness. \\ \\
Using the sate vector $x := (\theta, \dot{\theta}, e_a)$ and the control input $u = i_a$, the ODEs governing the dynamics can be written as follows:
\begin{figure}
\centering
\includegraphics[width=0.3\textwidth]{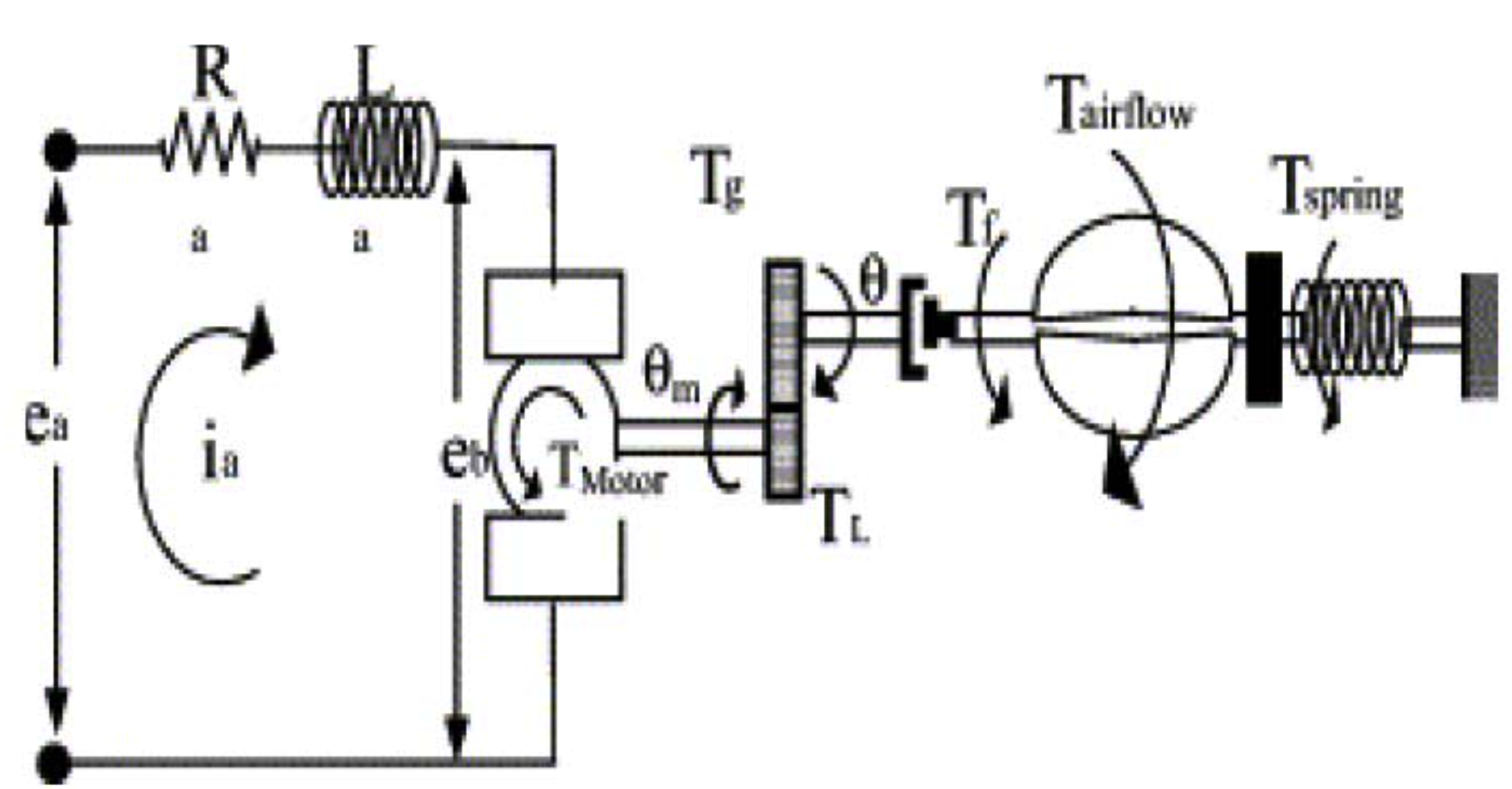} 
\caption{\textbf{Example 2:} Schematics of the automotive electronic throttle controlled system.}\label{fig_etc}
\end{figure}
\begin{subequations}\label{eqetc}
\begin{align}
\dot x_1=&x_2 \\
\dot x_2=&\dfrac{1}{J}\bigl[-K_{sp}(x_1-\theta_0)-K_fx_2+NK_tx_3\\ 
& -\pi R_p^2R_{af}\Delta_p\cos^2(x_1)\bigr]\\
\dot x_3=&\dfrac{1}{L_a}\bigl[-NK_bx_2-R_ax_3+u\bigr]
\end{align}
\end{subequations} 
\noindent where $K_{sp}$ characterizes the spring stiffness (see Figure \ref{fig_etc}), $K_f$ is the friction coefficient, $N$ is the gear ratio, $K_t$ is the electric torque static gain relatively to the current $i_a$. $R_p$ i s a constant radius, $\Delta_p = P_{atm} - P_m$ where $P_m = f(\theta, P_{atm}, N)$ is the manifold pressure that approaches the atmospheric pressure $P_{atm}$ when the throttle is wide open. 
\begin{rmk}
It is important to recall that we are interest in the situations where \textbf{none} of these facts regarding the equations governing the system is available to the designer of anomaly detectors.  \hfill $\spadesuit$
\end{rmk}
The nominal values of the parameters involved in the dynamics (\ref{eqetc}) are displayed in the table below:
\begin{center}
\begin{minipage}{0.35\textwidth}
\begin{table}[H]
\begin{tabular}{llllll}
\hline
\multicolumn{3}{c}{\textbf{Nominal parameters in \eqref{eqetc}}} \\
\hline
$K_{sp}$ & 0.4316     & $R_p$     & 0.0015   \\
$K_f$    & 0.4834     & $R_{af}$     & 0.002   \\
$N$     & 4  & $K_b$     & 0.1051   \\
$K_t$    & 0.1045   & $L_a$   & 0.003   \\
\hline
\end{tabular}
\end{table}	
\end{minipage} 
\end{center} 
Recall that the aim of the controller is to force the angle $\theta$ to follow some reference trajectory $\theta_{ref}$, which is given by an exogenous control law that manipulates the quantity of air entering the system. While it is obviously beyond the scope of this paper to go into the details of the controller design, we can nonetheless mention that the implemented feedback takes the following form:
\begin{equation}
u(t) = F_b\Bigl(\vec{\theta}(t), \vec{i_a}(t), \vec{u}(t), \{\vec{\theta}_\text{ref}^{(i)}(t)\}_{i=0,1,2}\Bigr)
\end{equation} 
where $\vec{\theta}(t)$, $\vec{i_a}(t)$ and $\vec{u}(t)$ stand for the measurement profiles over some past time window, while $\theta_\text{ref}^{(i)}(t)$ stands for the $i$-th derivative of the reference profile. \\
These inputs enable the reconstruction of the whole state to be used in some state feedback control expression which is, in our case, based on the back-stepping design. Namely, the knowledge of $\theta_{ref}$ is used to define a reference on $x_3$ (viewed as control for $x_2$), which induces a control design on $u$. The details are skipped here for the sake of brevity and since it lies outside the scope of the present paper. \\ \\ 
Regardless of the mechanisms behind the produced time-series, we assume that one disposes of recording of the two sensors defined by:
\begin{equation}
    s := [x_1, u]
\end{equation}
Behind the scene and for the sake of clarity, these time-series have been produced by simulating the closed-loop system for a sequence of step-changes following the same logic as the one used previously in Example 1 (Section \ref{subsec_lorentz}). \e
The sampling time is taken equal to $\tau = 0.02$s. Six closed-loop simulations (of 1 hour each) have been executed. Among them, three used the nominal values of the parameters $K_b$, $K_t$ and $L_a$, while in each of the three remaining simulations one of these parameters has been changed by 30\%. \\ \\
\begin{figure}
\centering
\begin{minipage}{.45\textwidth}
  \centering
  \includegraphics[width=0.9\textwidth]{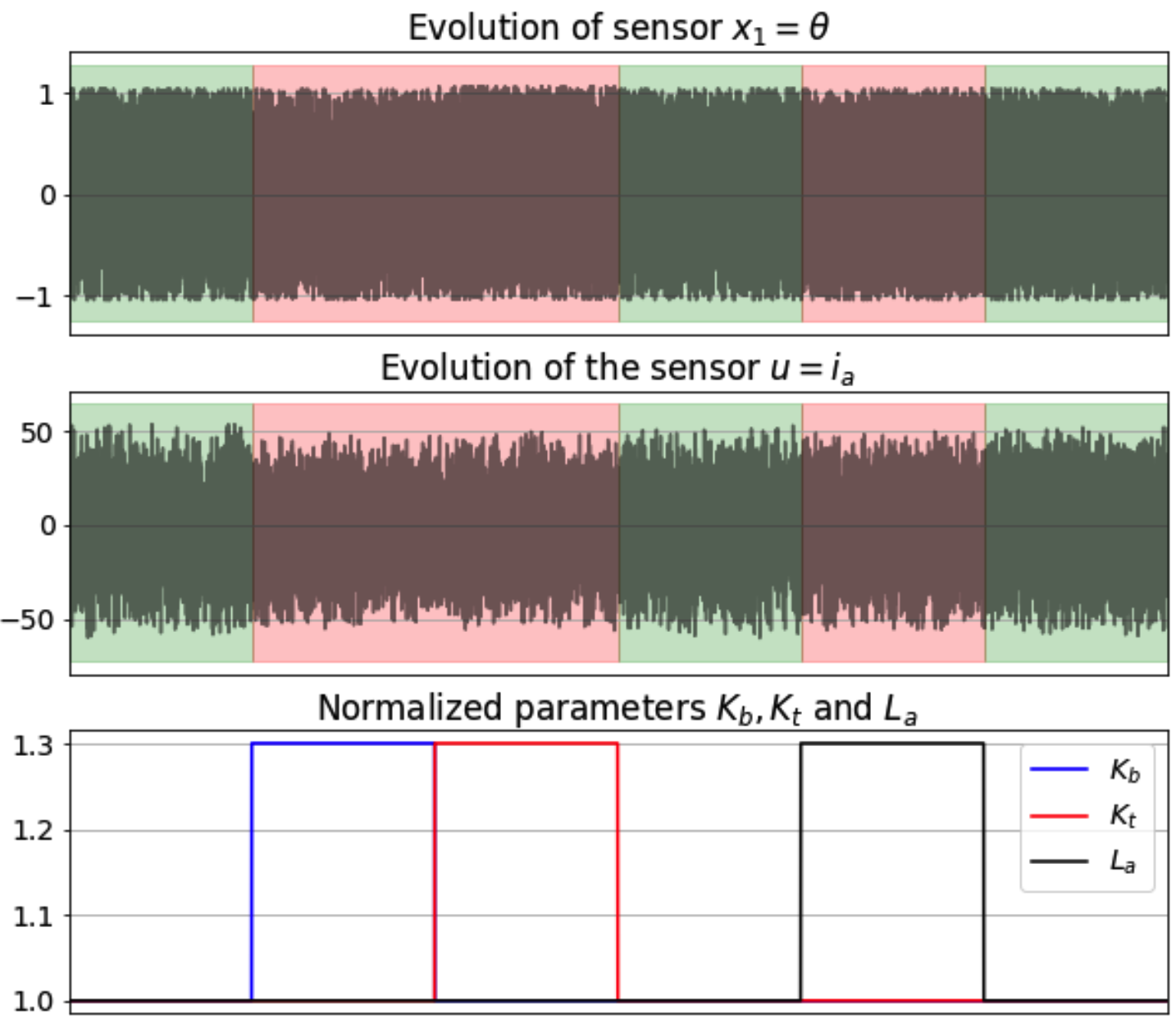} 
  \caption{\textbf{Example 2:} Raw time-series of $\theta$ and $u$ in the working data. The first nominal green region is used for learning the anomaly detector.}\label{fig:raw_etc}  	
\end{minipage}%
\hspace{0.05\textwidth}
\begin{minipage}{.45\textwidth}
  \centering
  \includegraphics[width=0.9\textwidth]{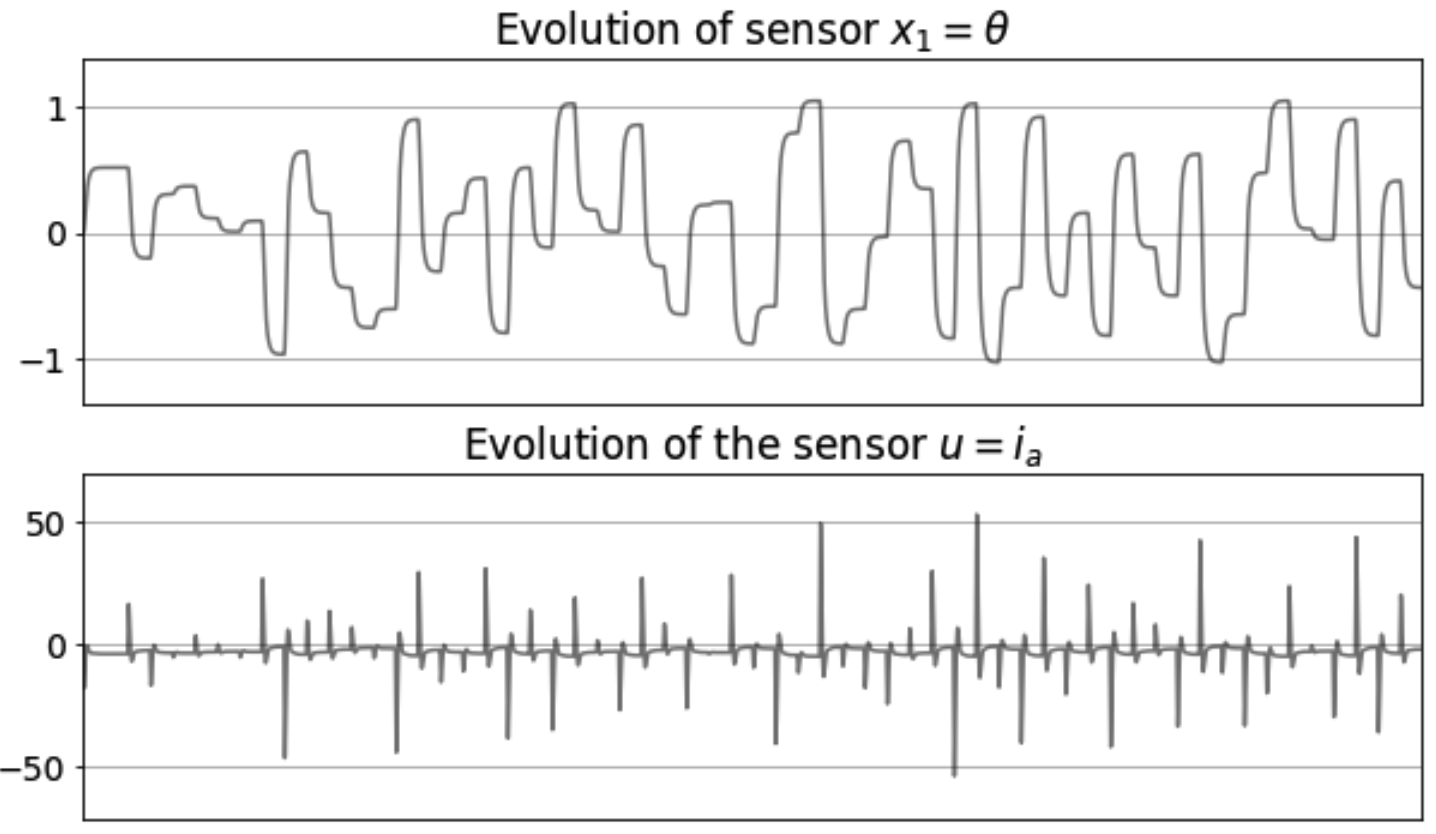} 
  \caption{\textbf{Example 2:} zoom on the raw time-series of Figure \ref{fig:raw_etc} showing the sequence of successive step changes in the reference $\theta$ and the associated control input profile.}\label{fig:raw_etc_zoom}	
\end{minipage}
\end{figure}
\noindent The raw time-series of sensors $x_1 = \theta$ and $u = i_a$ are shown in Figure \ref{fig:raw_etc}. To illustrate the influence of the control input sensor $u = i_a$ on the reference sensor $x_1 = \theta$ and the subsequent step changes induced, a zoom on the raw time-series is displayed on Figure \ref{fig:raw_etc_zoom}. \\ \\
Here again, the \textbf{green regions} correspond to intervals where the nominal values of the parameters are used although with different sequences of reference step changes. They represent a \textbf{normal behavior}. \\ \\
The \textbf{red regions} (second, third and fifth intervals) correspond to different different detuned sets of parameters as shown in the last row of plots. They represent \textbf{faulty behaviors}. \\ \\
The anomaly detection hyper-parameters are taken as follows:
\begin{equation}
n_q=10,\quad  \Delta=20, \quad \eta=0.95 \quad \epsilon = 0\label{etcparam}
\end{equation} 
The various evolutions of the family of residuals representing the point of view of sensors $x_1 = \theta$ and $u=i_a$ are displayed in Figure \ref{fig:etc_results_1}.

\begin{rmk}[on the explainability]\ \\
It can be noticed that the transition-related residual $r_{\textrm{trans}}$ does not bring any information regarding the presence of anomalies in this example as no anomalous transition have been found. On the other hand, as expected given the closed-loop nature of the example (where the controller compensates for the parameters changes so that the angle $\theta$ roughly follows the desired reference), the bounds-related residual $r_{\textrm{bound}}$ viewed by sensor $x_1 = \theta$ is barely useful as the excursions of $\theta$ are rightly controlled by the feedback law. \\ \\
This does not apply to the configuration-related residual $r_{\textrm{conf}}$ viewed by this sensor ($x_1$), since this residual involves also the values of the control being applied to force the desired behavior. \\ \\
It remains however obvious that the contributions of the residuals $r_{\textrm{bound}}$ and $r_{\textrm{conf}}$ viewed by the sensor $u=i_a$ are far more important in the anomaly detection. Notice also how $r_{\textrm{bound}}$ fails in detecting the variation on the parameter $L_a$ while this change is clearly visible on the residual $r_{\textrm{conf}}$ viewed by $u$. 
\hfill $\spadesuit$ 
\end{rmk}
\section{Comparison to common used algorithms}\label{sec_bench}
\noindent In ordert to evaluate the performances of the introduced method, four common Anomaly Detection algorithms have been selected to be tested on the two previous datasets. 
\e In order to compare the different solutions and since the labeling is perfect, usual Anomaly Detection metrics are proposed, such as the AUC-ROC which focuses on the trade-off between recall and false positive, the pAUC-ROC which is the same metric but with a maximum false positive rate of 10\% and the F1-score which provides a trade-off between precision and recall. Let us first recall the principle of the different alternatives and give some information regarding the associated implementation. 

\subsection{Algorithms implementations}\label{subsec_algo_description}
\noindent Each algorithm has been implemented to run on windows of 100 points. To go through the diversity of parameters, each method has been computed multiple times using various parameterization sets, and the evaluation metrics of the results have been averaged. As some algorithms computing time explodes with bigger time-series, there can be a difference of number of iterations between the datasets. \\
\subsubsection{Linear Regression - Lasso LARS}
\noindent The Linear Regression is one of the oldest algorithms for signal reconstruction, and has found applications in Anomaly Detection \cite{Chabi2023}. For this application, multiple sub-time-frame (of size $\leq 100$) have been estimated, then for each the residual is coined as the reconstruction error between a sensor and its estimation using its past values in the sub-time-frame and all the values from the sub-time-frame of the other sensors, then the results are smoothed using the maximum of a sliding time-frame of size 100. The implementation is done using \texttt{Scikit-Learn} package  \cite{Pedregosa2011sklearn}, using the Lasso LARS (Least Angle Regression) \cite{Efron2004} optimization. The algorithm has been run on both sensors, varying the optimization coefficient for each iteration. There is a total of 2600 runs for each dataset. \\
\subsubsection{LOF}
\noindent The LOF \cite{Breunig2000}, or Local Outlier Factor, is a non time-aware Anomaly Detection algorithm. Its residual is based on the point local deviation in regard to the density of its neighborhood. For this application, a multivariate point is coined as the concatenation of a sub-time-frame of each sensor (of size $\leq 100$), then the results are smoothed using the maximum of a sliding time-frame of size 100. The integration chosen is the \texttt{PyOD} one \cite{Zhao2019pyod}, and the algorithm has been run 27 times for each dataset, varying the neighborhood size and the nearest neighbors method for each iteration. \\
\subsubsection{Random Forest}
\noindent The Random Forest is an algorithm based on the training of numerous decision trees using various sub-samples of the dataset, outputting the average result. Here again, multiple sub-time-frame (of size $\leq 100$) have been estimated, then for each the residual is coined as the reconstruction error between a sensor and its estimation using its past values in the sub-time-frame and all the values from the sub-time-frame of the other sensors, then the results are smoothed using the maximum of a sliding time-frame of size 100. The integration chosen is the \texttt{Scikit-Learn} one \cite{Pedregosa2011sklearn}, and the algorithm has been run on both sensors, varying the trees configurations for each iteration. There is a total of 135 runs for the Lorentz Attractor dataset and 30 for the ETC dataset. \\
\subsubsection{Auto-Encoder}
\noindent As already sketched in Section \ref{subsec_dnn}, the  Auto-Encoder is a type of neural networks that consists in two parts: the first part serves an encoding function, reducing the data dimension, and the second serves a decoding function, reconstructing the data from its reduced form. For this application, it has been decided that the residual is defined to be the reconstruction error between the current values of both sensors and the estimation using their past values in a sub-time-frame (of size $\leq 100$), then the results are smoothed using the maximum of a sliding time-frame of size 100. The integration chosen is the \texttt{PyOD} one \cite{Zhao2019pyod}. In order to explore various hidden layers size and number, as well as the activation functions, the algorithm has been run 24 times. 
\subsection{Proposed algorithm}\label{subsec_bench_proposed}
\noindent For this application, the proposed algorithm has been run using 100-points windows. The configuration-related
residuals as well as the bounds-related residuals has been normalized using their values on the training dataset, and
the transition-related residuals have not been averaged on the windows: with these implementations, the final output
outlier score is the maximum of the three residuals. To explore a large combination of the algorithm parameters, it has been run 560 times for
the Lorentz Attractor dataset, and 138 times for the ETC dataset. In order to emulate an approach without any prior knowledge, the rules of thumb coined in \ref{subsec_hyperpameters_selection} haven't been used in this simulation, and the parameters range from within to well outside the advised values spans. 
\begin{table*}
\centering
\begin{tabular}{|c||c|c|c|}
     \multicolumn{4}{c}{Lorentz Attractor}  \\
     \hline 
     Metric & ROC-AUC ± std (max) & pAUC ± std (max) & F1 ± std (max) \\
     \hline
     LassoLars & 0.500 ± 0.016 (0.547) & 0.499 ± 0.003 (0.509)	& 0.153 ± 0.014 (0.220) \\
     LOF & 0.620 ± 0.088 (0.767)	& 0.517 ± 0.015 (0.556) & 0.261 ± 0.063 (0.467)	 \\
     RandomForest & \textit{0.742} ± 0.031 (\textit{0.835}) & \textit{0.539} ± 0.016 (\textit{0.569}) & \textit{0.377} ± 0.092 (\textit{0.554})  \\
     AutoEncoder & 0.550 ± 0.042 (0.646)	 & 0.508 ± 0.031 (0.517) & 0.203 ± 0.031 (0.274)  \\
     Proposed algorithm& \textbf{0.850} ± 0.150 (\textbf{0.997}) & \textbf{0.765} ± 0.150 (\textbf{0.986}) & \textbf{0.693} ± 0.230 (\textbf{0.981})  \\
     \hline 
     \multicolumn{4}{c}{ETC}  \\
     \hline 
     Metric & ROC-AUC ± std (max) & pAUC ± std (max) & F1 ± std (max)  \\
     \hline
     LassoLars & 0.522 ± 0.039 (0.823) & 0.509 ± 0.024 (\textit{0.803}) & 0.190 ± 0.365 (0.682) \\
     LOF & \textbf{0.831} ± 0.061 (\textbf{0.922}) & \textit{0.575} ± 0.005 (0.648)	 &  \textit{0.430} ± 0.135 (\textit{0.707})  \\
     RandomForest & 0.550 ± 0.011 (0.558) & 0.500 ± 0.050 (0.513) &  0.165 ± 0.022 (0.216)  \\
     AutoEncoder & 0.502 ± 0.023 (0.553)	& 0.496 ± 0.007 (0.515) & 0.148 ± 0.033 (0.222)	 \\
     Proposed algorithm & \textit{0.702} ± 0.133 (\textit{0.882}) & \textbf{0.612} ± 0.102 (\textbf{0.808}) & \textbf{0.462} ± 0.243 (\textbf{0.828})  \\
     \hline
\end{tabular}
\caption{\textbf{Benchmark results:} The metric values displayed are the average of all runs of each algorithms, with the standard deviation and the maximum. For the average and the maximum, the best results are in bold, second best in italics. The ROC-AUC stands for the standard ROC-AUC metric, the pAUC for the ROC-AUC metric but computed only on the thresholds that allow for a False Positive Rate inferior to 10\%, and the F1 for the standard F1-score.}\label{table:bench}
\end{table*}
\begin{table*}
\centering
\begin{tabular}{|c||c|c|c|c|c|c|}
     \multicolumn{7}{c}{Lorentz Attractor}  \\
     \multicolumn{1}{c}{}& \multicolumn{2}{c}{Smoothing window = 500} & \multicolumn{2}{c}{Smoothing window = 1000} & \multicolumn{2}{c}{Smoothing window = 5000} \\
    \hline
     Metric & ROC-AUC (max) & pAUC (max)& AUC (max) & pAUC (max)& AUC (max) & pAUC (max)\\
     \hline
     LassoLars & 0.506 (0.562) & 0.467 (0.570) & 0.512 (0.586) & 0.455 (0.585) & 0.537 (0.685) & 0.462 (0.679) \\
     LOF & 0.518 (0.532) & 0.581 (0.716) & 0.522 (0.542) & 0.562 (0.679) & 0.594 (0.695) & 0.592 (0.656)  \\
     RandomForest & 0.511 (0.537) & 0.685 (0.762) & 0.504 (0.528) & 0.668 (0.766) & 0.506 (0.600) & 0.563 (0.654) \\
     AutoEncoder & 0.522 (0.602) & 0.511 (0.548) & 0.522 (0.601) & 0.506 (0.553) & 0.511 (0.642) & 0.555 (0.625) \\
     Proposed algorithm & \textbf{0.859} (\textbf{0.995}) & \textbf{0.754} (\textbf{0.976}) & \textit{0.856} (\textit{0.991}) & \textit{0.739} (\textit{0.962}) & \textit{0.810} (\textit{0.988}) & \textit{0.703} (\textit{0.947}) \\
     \hline 
     \multicolumn{7}{c}{ETC}  \\
     \multicolumn{1}{c}{}& \multicolumn{2}{c}{Smoothing window = 500} & \multicolumn{2}{c}{Smoothing window = 1000} & \multicolumn{2}{c}{Smoothing window = 5000} \\
    \hline
     Metric & ROC-AUC (max) & pAUC (max)& AUC (max) & pAUC (max)& AUC (max) & pAUC (max)\\
     \hline
     LassoLars & 0.515 (0.833) & 0.553 (0.837) & 0.517 (0.831) & 0.558 (0.832) & 0.527 (0.836) & 0.525 (0.818)\\
     LOF & 0.565 (0.637) & \textbf{0.769} (0.864) & 0.570 (0.649) & \textit{0.745} (0.848) & 0.593 (0.750) & \textit{0.710} (0.849) \\
     RandomForest & 0.500 (0.529) & 0.530 (0.562) & 0.500 (0.537) & 0.522 (0.574) & 0.507 (0.600) & 0.483 (0.602)\\
     AutoEncoder & 0.518 (0.577) & 0.517 (0.534) & 0.539 (0.614) & 0.522 (0.540) & 0.630 (0.756) & 0.532 (0.559) \\
     Proposed algorithm& \textit{0.776} (\textit{0.972}) & 0.657 (\textit{0.907}) & \textit{0.789} (\textit{0.981}) & 0.672 (\textit{0.941}) & \textbf{0.795} (\textbf{0.995}) & 0.689 (\textbf{0.977})\\
     \hline 
\end{tabular}
\caption{\textbf{Benchmark results with smoothing:} The metric values displayed are the average of all runs of each algorithms, according to the smoothing in header. The output score of each algorithms have been smoothed using the maximum value of a sliding window. Overall best results are in bold, best for the smoothing are in italics.}\label{table:bench_smoothing}

\end{table*}
\subsection{Results}\label{subsec_results}
\noindent The results are displayed in Table \ref{table:bench} which suggests that while the proposed algorithm can be punctually outperformed by other algorithms (by the LOF using the ROC-AUC for the ETC dataset), it has overall both the best results in a blind uses, as seen with its average performances, and in tuned uses, as seen with its max performances. Other algorithms display interesting results with a blind approach, such as the LOF and the RandomForest for the Lorentz Attractor dataset or the LOF for the ETC dataset. While not performing with a blind approach, the LassoLars shows high max performance for the ETC dataset. \e
A second bench has been constructed in Table \ref{table:bench_smoothing}, simulating the industrial habit of aggregating the results on a temporal window: for windows of varying size, all residuals in this segment have been aggregated as the maximum value. While the proposed algorithm doesn't display significant performance deviations, some models such as the LOF or the Auto-Encoder do. 
\begin{rmk}[on the high variance]
One can notice a higher variance for the proposed solution than for the others. It is due to the hyper-parameters influence mentioned in \ref{subsec_hyperpameters_selection}, as hyper-parameters out of the rules of thumb ranges can reduce the algorithm performance. Nevertheless, as we aim for a blind approach for every evaluated model, we decided to not exclude theses hyper-parameter values. \hfill $\spadesuit$
\end{rmk}
\section{Conclusion and future works}\label{sec_conclusion}
\noindent In this contribution, a new \textit{conventional} (as defined in Section \ref{subsec_conventional}) method is proposed for anomaly detection in industrial time-series. The method is based on a clusterization of the data using the sensors transitions in order to define, over each sensor, three residuals representing potentially complementary distances to normality. The proposed residuals are not based on any Machine Learning step and enable a high level of explainability. Furthermore, they display a high potential for compatibility with Operator Feedback which is unavoidable in the industrial context under interest, where the transient and context-dependent consistence of labels are not necessary agreeable with standard ML metrics. \\
Two use cases are proposed in order to show the efficiency of the proposed methodology and the complementarity between the different residuals (as illustrated in Remark \ref{rmk_complementarity}) in both cyclic and non cyclic contexts. \\ \\
Although the previous results suggest some relevance of the proposed algorithm, future works should include more thorough benchmarks with industrially-gathered data, and broaden the focus around the algorithm explainability evaluation while taking in account the computing time and memory usage. Moreover, the feedback friendliness claim (see Sections \ref{subsec_labelling} and \ref{subsec_np3}) should be highlighted and rigorously tested. \\ \\
In spite of the encouraging results as assessed by the proposed use-cases, it is a fact that the present paper intentionally only skimmed a fundamental step which is crucial for the complete design of agnostic anomaly detector, namely the computation of threshold and the merging of the different residuals into a single one that might be used in the final almost binary decision making step. Nevertheless, to match the benchmarking needs, a simple aggregation method has been briefly proposed in Section \ref{sec_bench}. \\ \\
This choice is not only related to the size of the paper, but reflects our deep belief that in the industrial time-series case, this concept of binary decision with a single residual (refer to the discussion provided on the specificities of the situation in Section \ref{subsec_discussions}) is not the appropriate one and, therefore, what is needed is not simply some choice here and there to come out with a standard binary counting-based end-process decision. Rather, a whole reformulation of what is really expected by operators and what would be really compatible with the manners with which general defaults show themselves in industrial time-series need to be seriously undertaken. 
\bibliography{main}
\bibliographystyle{plain}
\appendix 
\section{On controlling the cardinality of the uncorrelated sets of configurations}\label{appKmeans}
\noindent We propose to use a \textbf{K-Means} \cite{MacQueen1967} algorithm on the uncorrelated sensors' configurations sets with a cutoff number of clusters $n_w$. In this case, the third set of normality parameters for a sensor $i$ would be: 
\begin{equation}
\centering
\begin{aligned}
\mathbf{NP}_3^{[i]}({\mathcal{T}_{train}}) := \Bigl\{\textbf{K-Means}(\widetilde{\mathcal{W}}^{[i]}_{c}(\mathcal{T}_{train}) \text{, } \\ \text{n\_clusters}=n_w) \text{ \(|\) } c \in c^{[i]}_{\mathcal{T}_{train}} \Bigr\} \label{def_NP3_ith_kmeans}
\end{aligned}
\end{equation}
And an upper bound of this NP set cardinality can be expressed as follows:
\begin{equation}
\centering
\begin{aligned}
|\mathbf{NP}_3({\mathcal{T}_{train}})| \leq \underset{\begin{subarray}{c}
  \text{config. vectors} \\
  \text{length}
  \end{subarray}}{\uwave{(\Delta + N_s - 1)}} \cdot n_w \cdot \sum^{N_s}_{i=1}|c^{[i]}_{\mathcal{T}_{train}}| \\ \leq (\Delta + N_s - 1) \cdot n_w \cdot N_s \cdot {n_q}^2 \label{def_np3_cardinality}
\end{aligned}
\end{equation}
\section{On the hyper-parameters selection}\label{subsec_hyperpameters_selection}
\begin{figure*}[ht]
    \centering
    \includegraphics[scale=0.45]{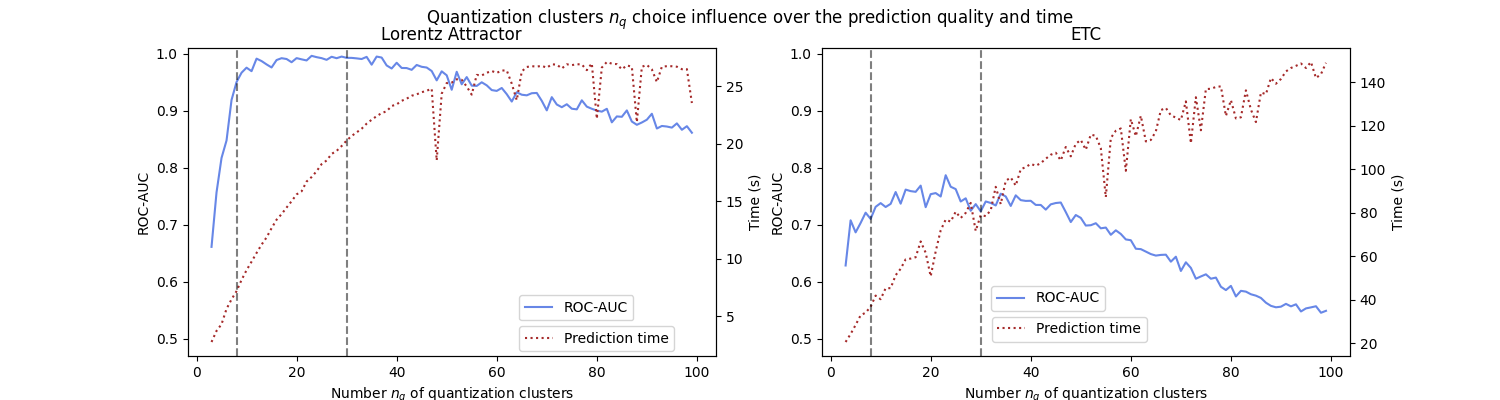}
    \caption{Influence of the selection of the number $n_q$ of quantization clusters over the prediction quality and time, evaluated on the two example datasets. The performance metric chosen is the ROC-AUC. Notice the area between dashed line, which represents the range of $n_q$ values mentioned in the rule of thumb above. The hyper-parameters, excepting $n_q$, are $\Delta = 20$, $\eta = 0.95$ and $\epsilon = 1$.}
    \label{fig:influence_nq}
\end{figure*}
\begin{figure*}[ht]
    \centering
    \includegraphics[scale=0.45]{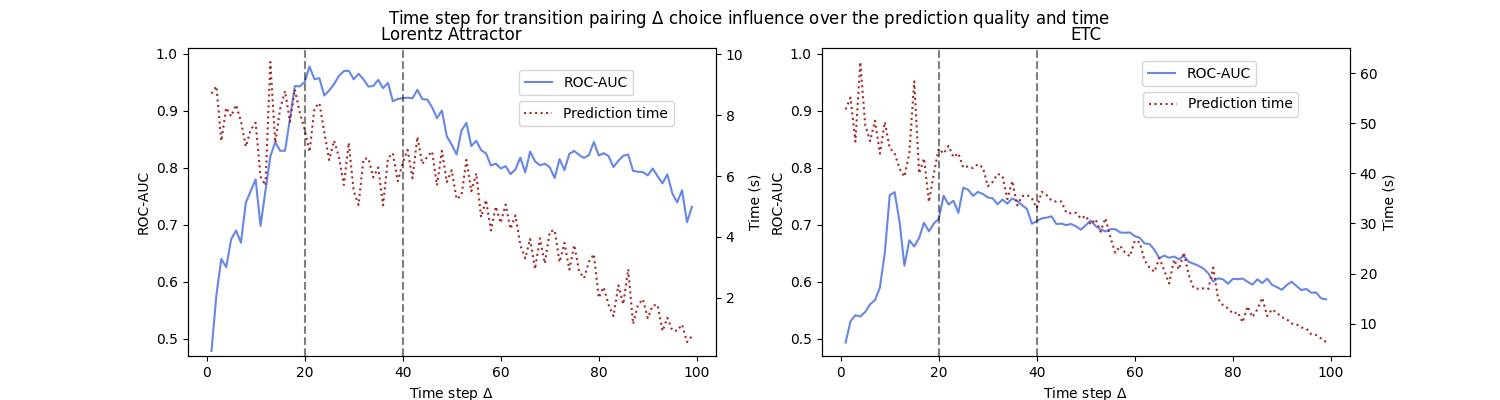}
    \caption{Influence of the selection of the number $\Delta$ of time steps for transition pairing over the prediction quality and time, evaluated on the two example datasets. The performance metric chosen is the ROC-AUC. Notice the area between dashed line, which represents the range of $\Delta$ values mentioned in the rule of thumb above. The hyper-parameters, excepting $\Delta$, are $n_q = 8$, $\eta = 0.95$ and $\epsilon = 1$.}
    \label{fig:influence_delta}
\end{figure*}
\noindent As a hyper-parameters tuning module has yet to be successfully added to the proposed framework, there is no way to perfectly estimate the hyper-parameters without prior knowledge on the system. Nevertheless, through extensive evaluations on various synthetic and real datasets, some \textit{rules of thumb} have been coined as general guidelines. These rules are proposed bellow, illustrated using the two realistic datasets proposed in Section \ref{subsec_lorentz} and \ref{subsec_etc}.\\ \\
Figure \ref{fig:influence_nq} shows that an acceptable trade-off between performance and computing time can be found for a number of \textbf{quantization clusters} ranging from 8 to 30. As this hyper-parameter depends of the system dynamics but also on the ratio between values range and noise, it needs to be tuned accordingly to the values granularity. \\ \\
Figure \ref{fig:influence_delta} shows that there also exists such kind of trade-off for the \textbf{time step for transition pairing}, and range from 20 to 40. As this hyper-parameter heavily depends of the sensor nature and sampling frequency, it is to be tuned accordingly. Notice the computing time decreasing while $\Delta$ is increasing, which is only true if there is no sliding window used, resulting in a higher $\Delta$ meaning less transitions analyzed overall.
\section{Incremental Learning}
\subsection{Increasing the Normality Characterizations}\label{il_increase_norm}
\noindent For a sub-time-frame $\mathcal{T}_{new}$ of the mislabeled points and a training time interval $\mathcal{T}_{train}$, the \textbf{Normality Parameters}
sets are updated for each sensor $i$ as follows: 
\begin{center}
\begin{minipage}{0.46\textwidth} 
\begin{equation}
\centering
\begin{split}
  \mathbf{NP}^{[i]}_1({\mathcal{T}_{train} + \mathcal{T}_{new}}) := & \mathbf{NP}^{[i]}_1({\mathcal{T}_{train}})
  \\ & \cup \Bigl\{c \in c^{[i]}_{\mathcal{T}_{new}} \text{ \(|\) } c \notin c^{[i]}_{\mathcal{T}_{train}} \Bigr\} \label{def_np1_IL_increase}
\end{split}
\end{equation}

\end{minipage}
\end{center}

\begin{center}
\begin{minipage}{0.46\textwidth} 
\begin{equation}
\centering
\begin{split}
  \mathbf{NP}_2^{[i]}({\mathcal{T}_{train} + \mathcal{T}_{new}}) &:= \Bigl\{(\underline{b^{[i]}_c}(\mathcal{T}_{train} + \mathcal{T}_{new}), 
  \\ & \overline{b^{[i]}_c}(\mathcal{T}_{train} + \mathcal{T}_{new})) \text{ \(|\) } c \in c^{[i]}_{\mathcal{T}_{train}} \Bigr\}
  \\ & \cup \Bigl\{(\underline{b^{[i]}_c}(\mathcal{T}_{new}), \overline{b^{[i]}_c}(\mathcal{T}_{new})) 
  \\ & \text{ \(|\) } c \in c^{[i]}_{\mathcal{T}_{new}}, c \notin c^{[i]}_{\mathcal{T}_{train}} \Bigr\} \quad \textrm{where} \label{def_np2_IL_increase}
\end{split}
\end{equation}

\end{minipage} 
\end{center}

\begin{equation}
\begin{split}
    \underline{b^{[i]}_c}(\mathcal{T}_{train} + \mathcal{T}_{new}) := & \Bigl\{ \min \Bigl(\underline{b^{[i]}_c}(\mathcal{T}_{train}), \underline{b^{[i]}_c}(\mathcal{T}_{new}) \Bigr)
    \\ & \text{ \(|\) } c \in c^{[i]}_{\mathcal{T}_{train}} \cap c^{[i]}_{\mathcal{T}_{new}} \Bigr\} 
    \\ & \cup \Bigl\{\underline{b^{[i]}_c}(\mathcal{T}_{train})
    \\ & \text{ \(|\) } c \in c^{[i]}_{\mathcal{T}_{train}}, c \notin c^{[i]}_{\mathcal{T}_{new}} \Bigr\} \label{def_bprime_under}
\end{split} 
\end{equation}
\begin{equation}
\begin{split}
    \overline{b^{[i]}_c}(\mathcal{T}_{train} + \mathcal{T}_{new}) := & \Bigl\{ \max \Bigl(\overline{b^{[i]}_c}(\mathcal{T}_{train}), \overline{b^{[i]}_c}(\mathcal{T}_{new}) \Bigr)
    \\ & \text{ \(|\) } c \in c^{[i]}_{\mathcal{T}_{train}} \cap c^{[i]}_{\mathcal{T}_{new}} \Bigr\} 
    \\ & \cup \Bigl\{\overline{b^{[i]}_c}(\mathcal{T}_{train})
    \\ & \text{ \(|\) } c \in c^{[i]}_{\mathcal{T}_{train}}, c \notin c^{[i]}_{\mathcal{T}_{new}} \Bigr\} \label{def_bprime_over}
\end{split}
\end{equation}

\begin{center}
\begin{minipage}{0.46\textwidth} 
\begin{equation}
\begin{split}
    \mathbf{NP}_3^{[i]}({\mathcal{T}_{train} + \mathcal{T}_{new}}) := &\Bigl\{\widetilde{\mathcal{W}}^{[i]}_{c}(\mathcal{T}_{train}) \cup \widetilde{\mathcal{W}}^{[i]}_{c}(\mathcal{T}_{new})
    \\ & \text{ \(|\) } c \in c^{[i]}_{\mathcal{T}_{train}} \cap c^{[i]}_{\mathcal{T}_{new}} \Bigr\}
    \\ & \cup \Bigl\{\widetilde{\mathcal{W}}^{[i]}_{c}(\mathcal{T}_{train}) 
    \\ & \text{ \(|\) } c \in c^{[i]}_{\mathcal{T}_{train}}, c \notin c^{[i]}_{\mathcal{T}_{new}} \Bigr\}
    \\ & \cup \Bigl\{\widetilde{\mathcal{W}}^{[i]}_{c}(\mathcal{T}_{new}) 
    \\ & \text{ \(|\) } c \in c^{[i]}_{\mathcal{T}_{new}}, c \notin c^{[i]}_{\mathcal{T}_{train}} \Bigr\}\label{def_np3_IL_increase}
\end{split}
\end{equation}

\end{minipage} 
\end{center}
where the extremum operators mentioned in \ref{def_bprime_under} and \ref{def_bprime_over} are the point by point max and min of the concatenated vectors. 

\subsection{Reducing the Normality Characterizations}\label{il_reduce_norm}
For a sub-time-frame $\mathcal{T}_{new}$ of the undetected points, a training time interval $\mathcal{T}_{train}$ and a correlation tolerance factor $\zeta \in ]0,1[$, the \textbf{Normality Parameters}
sets are updated for each sensor $i$ as follows: 
\begin{center}
\begin{minipage}{0.46\textwidth} 
\begin{align}
\begin{split}
    \mathbf{NP}_3^{[i]}({\mathcal{T}_{train} - \mathcal{T}_{new}}) &:= \Bigl \{ \widetilde{\mathcal{W}}^{[i]}_{c}(\mathcal{T}_{train} - \mathcal{T}_{new}) 
    \\ & \\ & \text{ \(|\) } c \in c^{[i]}_{\mathcal{T}_{train}} \Bigr\} \quad \textrm{where}
\end{split} \label{def_np3_IL_reduce}
\end{align}

\end{minipage} 
\end{center}
\begin{equation}
\begin{split}
    \\  \widetilde{\mathcal{W}}^{[i]}_{c}(\mathcal{T}_{train} - \mathcal{T}_{new}) &:=   \Bigl\{ w \in \widetilde{\mathcal{W}}^{[i]}_{c}(\mathcal{T}_{train}) 
    \\ & \text{ \(|\) } \Bigl|\mathbf{corr}(w,  \widetilde{\mathcal{W}}^{[i]}_{c}(\mathcal{T}_{new})\Bigr| < \zeta \Bigr\} 
    \label{def_np3_IL_reduce_subdef}
\end{split}
\end{equation}

\begin{center}
\begin{minipage}{0.46\textwidth} 
\begin{equation}
    \begin{split}
    \mathbf{NP}_2^{[i]}({\mathcal{T}_{train} - \mathcal{T}_{new}}) :=&\Bigl\{(\underline{b^{[i]}_c}(\mathcal{T}_{train} - \mathcal{T}_{new}), 
    \\ & \overline{b^{[i]}_c}(\mathcal{T}_{train} - \mathcal{T}_{new}))
    \\ & \text{ \(|\) } c \in c^{[i]}_{\mathcal{T}_{train}} \Bigr\} \quad \textrm{where} \label{def_NP2_ith_reduce} 
    \end{split}
\end{equation}
\end{minipage} 
\end{center}

\begin{equation}
    \begin{split}
    \underline{b^{[i]}_c}(\mathcal{T}_{train} - \mathcal{T}_{new}) & := \\ & \bigl[\underset{w \in \widetilde{\mathcal{W}}^{[i]}_{c}(\mathcal{T}_{train} - \mathcal{T}_{new})}{min} w_j \bigr]_{1 \leq j \leq \Delta + N_s} \label{def_lower_bounds_reduce}
    \end{split}
\end{equation}
\begin{equation}
    \begin{split}
    \overline{b^{[i]}_c}(\mathcal{T}_{train} - \mathcal{T}_{new}) & := \\ & \bigl[\underset{w \in \widetilde{\mathcal{W}}^{[i]}_{c}(\mathcal{T}_{train} - \mathcal{T}_{new})}{max} w_j \bigr]_{1 \leq j \leq \Delta + N_s}\label{def_upper_bounds_reduce}
    \end{split}
\end{equation}

\begin{center}
\begin{minipage}{0.46\textwidth} 
\begin{equation}
\centering
\begin{split}
  \mathbf{NP}^{[i]}_1({\mathcal{T}_{train} - \mathcal{T}_{new}}) := &  \Bigl\{c \in c^{[i]}_{\mathcal{T}_{train}}
  \\ & \text{ \(|\) } \widetilde{\mathcal{W}}^{[i]}_{c}(\mathcal{T}_{train} - \mathcal{T}_{new}) \neq \emptyset \Bigr\} \label{def_np1_IL_reduce}
\end{split}
\end{equation}

\end{minipage} 
\end{center}
where the $\mathbf{corr}$ operator is the one defined in \ref{def_correlation}.

\section{Notations}\label{appendix_parameters}
\subsection{hyper-parameters}
\ \\
\begin{itemize}
    \item[] \begin{center}Fitting hyper-parameters \end{center}
    \begin{rightbrace}
    \item $\mathbb{D}_{train}$: Training dataset (see \ref{def_Dtrain})
    \item $n_q$: Number of quantization clusters (see \ref{def_interval}, \ref{def_quantizer},  \ref{def_np3_cardinality})
    \item $\Delta$: Time step for transition pairing (see \ref{def_transition}, \ref{def_configuration_vector}, \ref{def_lower_bounds}, \ref{def_upper_bounds}, \ref{def_np3_cardinality}, \ref{def_constraint_win_size}, \ref{def_bound_delta})
    \item $\eta$: Cutoff absolute correlation threshold for configuration vector sets reduction (see \ref{def_configuration_set_lowcorr})
    \item $n_w$: Cutoff cardinality for configuration vector sets clustering (see \ref{def_NP3_ith_kmeans}, \ref{def_np3_cardinality}) [Optional]
    \end{rightbrace}
    \item[] \begin{center}Prediction hyper-parameters \end{center}
    \begin{rightbrace}
    \item $n_{pred}$: Sampling length for prediction (see \ref{def_constraint_win_size}, \ref{def_res_trans}, \ref{def_res_bound})
    \item $\epsilon$: Bounds deviation tolerance for domain-related residual (see \ref{def_distance})
    \end{rightbrace}
\end{itemize}
\subsection{Model parameters}
\begin{itemize}
    \item $N_s$: Number of sensors in the given dataset (see \ref{def_Dtrain}, \ref{def_NP1}, \ref{def_lower_bounds}, \ref{def_upper_bounds}, \ref{def_NP2}, \ref{def_NP3}, \ref{def_np3_cardinality}, \ref{def_bound_delta}, \ref{def_lower_bounds_reduce}, \ref{def_upper_bounds_reduce})
    \item $\mathcal{T}_{train}$: Time intervals used for the training part (see \ref{def_Dtrain}, \ref{def_quantization_cluster}, \ref{def_transition_set}, \ref{def_configuration_set}, \ref{def_configuration_set_timestamps}, \ref{def_NP1}, \ref{def_lower_bounds}, \ref{def_upper_bounds}, \ref{def_NP2}, \ref{def_NP2_ith},  \ref{def_configuration_set_lowcorr}, \ref{def_correlation}, \ref{def_NP3}, \ref{def_NP3_ith}, \ref{def_NP3_ith_kmeans}, \ref{def_np3_cardinality}, \ref{def_res_conf}, \ref{def_np1_IL_increase}, \ref{def_np2_IL_increase}, \ref{def_bprime_under}, \ref{def_bprime_over}, \ref{def_np3_IL_increase}, \ref{def_np3_IL_reduce}, \ref{def_np3_IL_reduce_subdef}, \ref{def_NP2_ith_reduce}, \ref{def_lower_bounds_reduce}, \ref{def_upper_bounds_reduce}, \ref{def_np1_IL_reduce})
    \item $s^{[i]}_{t}$: Value at instant $t$ of the $i$-th sensor,
    \item $\tilde s^{[i]}_{t}$: Scaled value at instant $t$ of the $i$-th sensor (see \ref{def_quantization_cluster}, \ref{def_configuration_vector})
    \item $\overline{N_s}$: The list of the sensors in the given dataset, in a numeric form (see \ref{def_NP1}, \ref{def_NP2}, \ref{def_NP3})
    \item $v^{[i]}_k$: $k$-th $n_q$-quantile of the scaled $i$-th sensor training values (see \ref{def_interval})
    \item $\mathcal{Q}^{[i]}$: $n_q$-quantizer (see \ref{def_quantizer})
    \item $q^{[i]}_{t}$: Cluster at instant $t$ of the $i$-th sensor (see \ref{def_quantization_cluster})
    \item $c^{[i]}_{t}$: Transition pair at instant $t$ for the $i$-th sensor (see \ref{def_transition}
    \item $c^{[i]}_{\mathcal{T}}$: Ensemble of transition pairs for the $i$-th sensor given some $\mathcal{T}$ time intervals (see \ref{def_transition_set}, \ref{def_NP1}, \ref{def_NP2_ith}, \ref{def_NP3_ith}, \ref{def_NP3_ith_kmeans}, \ref{def_np3_cardinality}, \ref{def_res_trans}, \ref{def_res_conf}, \ref{def_np1_IL_increase}, \ref{def_np2_IL_increase}, \ref{def_np1_IL_reduce}, \ref{def_bprime_under}, \ref{def_bprime_over}, \ref{def_np3_IL_increase}, \ref{def_np3_IL_reduce}, \ref{def_NP2_ith_reduce})
    \item $w^{[i]}_{t}$: Configuration vector at instant $t$ for the $i$-th sensor (see \ref{def_configuration_vector}, \ref{def_configuration_set}, \ref{def_bound_delta})
    \item $w_j$: $j$-th component of a given configuration vector $w$ (see \ref{def_lower_bounds}, \ref{def_upper_bounds}, \ref{def_lower_bounds_reduce}, \ref{def_upper_bounds_reduce}, \ref{def_np1_IL_reduce})
    \item $\mathcal{W}^{[i]}_c (\mathcal{T})$: Configuration vector set for a visited transition $c$ and the $i$-th sensor given some $\mathcal{T}$ time intervals (see \ref{def_configuration_set}, \ref{def_lower_bounds}, \ref{def_upper_bounds}, \ref{def_configuration_set_lowcorr}, \ref{def_correlation}, \ref{def_NP3_ith}, \ref{def_NP3_ith_kmeans}, \ref{def_res_conf})
    \item $[\mathcal{T}]^{[i]}_{c}$: Time intervals associated to a visited transition $c$ and the $i$-th sensor given some $\mathcal{T}$ time intervals (see \ref{def_configuration_set}, \ref{def_configuration_set_timestamps})
    \item $\mathbf{NP}_1({\mathcal{T}_{train}})$: First set of Normality Parameters (see \ref{def_NP1})
    \item $\mathbf{NP}^{[i]}_1({\mathcal{T}_{train}})$: First set of Normality Parameters for the $i$-th sensor (see \ref{def_np1_IL_increase}, \ref{def_np1_IL_reduce})
    \item $\underline{b^{[i]}_c}$: Lower bounds of excursion associated to a visited transition $c$ for the $i$-th sensor (see \ref{def_lower_bounds}, \ref{def_NP2_ith}, \ref{def_bound_delta}, \ref{def_np2_IL_increase}, \ref{def_bprime_under}, \ref{def_NP2_ith_reduce}, \ref{def_lower_bounds_reduce})
    \item $\overline{b^{[i]}_c}$: Upper bounds of excursion associated to a visited transition $c$ for the $i$-th sensor  (see \ref{def_upper_bounds}, \ref{def_NP2_ith}, \ref{def_bound_delta}, \ref{def_np2_IL_increase}, \ref{def_bprime_over}, \ref{def_NP2_ith_reduce}, \ref{def_upper_bounds_reduce})
    \item $\mathbf{NP}_2({\mathcal{T}_{train}})$: Second set of Normality Parameters (see \ref{def_NP2})
    \item $\mathbf{NP}_2^{[i]}({\mathcal{T}_{train}})$: Second set of Normality Parameters for the $i$-th sensor (see \ref{def_NP2}, \ref{def_NP2_ith}, \ref{def_np2_IL_increase}, \ref{def_NP2_ith_reduce})
    \item $\widetilde{\mathcal{W}}^{[i]}_{c}(\mathcal{T})$: Uncorrelated configuration vector set for a visited transition $c$ and the $i$-th sensor given some $\mathcal{T}$ time intervals (see \ref{def_configuration_set_lowcorr}, \ref{def_NP3_ith}, \ref{def_np3_IL_increase}, \ref{def_np3_IL_reduce}, \ref{def_np3_IL_reduce_subdef}, \ref{def_lower_bounds_reduce}, \ref{def_upper_bounds_reduce}, \ref{def_np1_IL_reduce})
    \item $\textbf{corr}(w, \mathcal{W})$: Vector of the unitary correlations between a vector $w$ and a vector set $\mathcal{W}$ (see \ref{def_configuration_set_lowcorr}, \ref{def_correlation}, \ref{def_res_conf})
    \item $\mathbf{NP}_3({\mathcal{T}_{train}})$: Third set of Normality Parameters (see \ref{def_NP3})
    \item $\mathbf{NP}_3^{[i]}({\mathcal{T}_{train}})$: Third set of Normality Parameters for the $i$-th sensor (see \ref{def_NP3}, \ref{def_NP3_ith}, \ref{def_NP3_ith_kmeans}, \ref{def_np3_IL_increase}, \ref{def_np3_IL_reduce})
    \item $\mathcal{T}_{pred}$: Time interval used for the prediction part (see \ref{def_res_trans}, \ref{def_res_bound}, \ref{def_res_conf})
    \item $r_{\textrm{trans}}$: Transitions-related residual
    \item $r_{\textrm{trans}}^{[i]}(\mathcal{T})$: Transition-related residual for the $i$-th sensor computed on some $\mathcal{T}$ time interval (see \ref{def_res_trans})
    \item $r_{\textrm{bound}}$: Bounds-related residual
    \item $r_{\textrm{bound}}^{[i]}(\mathcal{T})$: Bounds-related residual for the $i$-th sensor computed on some $\mathcal{T}$ time interval (see \ref{def_res_bound})
    \item $\delta^{[i]}(t)$ Out-of-bounds estimation for a configuration vector a instant $t$ for the $i$-th sensor (see \ref{def_res_bound}, \ref{def_bound_delta})
    \item $r_{\textrm{conf}}$: Configuration-related residual
    \item $r_{\textrm{conf}}^{[i]}(\mathcal{T})$: Configuration-related residual for the $i$-th sensor computed on some $\mathcal{T}$ time interval (see \ref{def_res_conf})
\end{itemize}
\end{document}